\documentclass[pdflatex,sn-mathphys-num]{sn-jnl}% Math and Physical Sciences Numbered Reference Style
\usepackage{graphicx}%
\usepackage{multirow}%
\usepackage{amsmath,amssymb,amsfonts}%
\usepackage{amsthm}%
\usepackage{mathrsfs}%
\usepackage[title]{appendix}%
\usepackage{xcolor}%
\usepackage{textcomp}%
\usepackage{footmisc}%
\usepackage{booktabs}%
\usepackage{algorithm}%
\usepackage{algorithmicx}%
\usepackage{algpseudocode}%
\usepackage{listings}%
\usepackage{fontawesome5} % 提供 GitHub 和 Hugging Face 图标
\usepackage{hyperref}      % 提供超链接功能
\usepackage{fancyhdr}

\theoremstyle{thmstyleone}%
\theoremstyle{thmstyletwo}%

\theoremstyle{thmstylethree}%

\begin{document}

\title[]{Vib2Mol: from vibrational spectra to molecular structures—a unified deep learning framework}

%%=============================================================%%
%% GivenName	-> \fnm{Joergen W.}
%% Particle	-> \spfx{van der} -> surname prefix
%% FamilyName	-> \sur{Ploeg}
%% Suffix	-> \sfx{IV}
%% \author*[1,2]{\fnm{Joergen W.} \spfx{van der Ploeg} 
%%  \sfx{IV}}\email{iauthor@gmail.com}
%%=============================================================%%

\author[1,2]{Xinyu Lu}\email{xinyulu@stu.xmu.edu.cn}

\author*[1,4]{Hao Ma}\email{oaham@xmu.edu.cn}

\author[3]{Hui Li}\email{hui@xmu.edu.cn}

\author[4]{Jia Li}\email{lijia@stu.xmu.edu.cn}

\author[1]{Yi Rong}\email{rongyi0107@stu.xmu.edu.cn}

\author[2,5]{Yuqiang Li}\email{liyuqiang@pjlab.org.cn}

\author[2,6]{Tong Zhu}\email{tongzhu.work@gmail.com}

\author*[7]{Guokun Liu}\email{guokunliu@xmu.edu.cn}

\author*[1,2,4]{Bin Ren}\email{bren@xmu.edu.cn}

\affil*[1]{\orgdiv{College of Chemistry and Chemical Engineering}, \orgname{Xiamen University}, \orgaddress{\city{Xiamen}, \postcode{361005}, \state{Fujian}, \country{China}}}

\affil*[2]{\orgname{Shanghai Innovation Institute}, \orgaddress{\city{Shanghai}, \postcode{200030}, \country{China}}}

\affil[3]{\orgdiv{School of Informatics}, \orgname{Xiamen University}, \orgaddress{\city{Xiamen}, \postcode{361005}, \state{Fujian}, \country{China}}}

\affil[4]{\orgdiv{Institute of Artificial Intelligence}, \orgname{Xiamen University}, \orgaddress{\city{Xiamen}, \postcode{361005}, \state{Fujian}, \country{China}}}

\affil[5]{\orgname{Shanghai Artificial Intelligence Laboratory}, \orgaddress{\city{Shanghai}, \postcode{200000}, \country{China}}}

\affil[6]{\orgdiv{School of Chemistry and Molecular Engineering}, \orgname{East China Normal University}, \orgaddress{\city{Shanghai}, \postcode{200062}, \country{China}}}

\affil[7]{\orgdiv{College of the Environment and Ecology}, \orgname{Xiamen University}, \orgaddress{\city{Xiamen}, \postcode{361005}, \state{Fujian}, \country{China}}}

%%==================================%%
%% Sample for unstructured abstract %%
%%==================================%%

\abstract{There will be a paradigm shift in chemical and biological research, to be enabled by autonomous, closed-loop, real-time self-directed decision-making experimentation. Spectrum-to-structure correlation, which is to elucidate molecular structures with spectral information, is the core step in understanding the experimental results and to close the loop. However, current approaches usually divide the task into either database-dependent retrieval and database-independent generation and neglect the inherent complementarity between them. In this study, we proposed Vib2Mol, a unified deep learning framework designed to flexibly handle diverse spectrum-to-structure tasks according to the available prior knowledge by bridging the retrieval and generation. Empowered by our coarse-to-fine retrieval and generate-then-rerank strategies, Vib2Mol not only achieves state-of-the-art performance in analyzing theoretical Infrared and Raman spectra, but also outperform previous models on experimental data. Moreover, our model demonstrates promising capabilities in predicting reaction products and sequencing peptides, enabling vibrational spectroscopy a potential guide for autonomous scientific discovery workflows.}

%%================================%%
%% Sample for structured abstract %%
%%================================%%


\maketitle

% --- 修改后的图标链接部分 ---
\begin{center}
    \vspace{-1.5em} 
    \hypersetup{colorlinks=true, urlcolor=black} 

    % 使用 tabular 环境，l 代表左对齐
    \begin{tabular}{l}
        % GitHub 链接
        \href{https://github.com/X1nyuLu/vib2mol}{%
            \faGithub\ \small https://github.com/X1nyuLu/vib2mol%
        } \\ % 换行
        % Hugging Face 链接
        \href{https://huggingface.co/datasets/xinyulu/vibench}{%
            \faLink\ \small https://huggingface.co/datasets/xinyulu/vibench%
        }
    \end{tabular}
    \vspace{1em}
\end{center}

\section{Introduction}
With the rapid development of automated experimental design and execution\cite{burger2020mobile,dai2024autonomous}, it has become possible to explore potential chemical reactions and study complex life processes with a closed-loop workflow without human intervention. It may significantly accelerate material design and drug discovery. The key to automating such a closed-loop workflow is to design and execute the next experiment on the basis of the prior knowledge. However, this is particularly challenging due to the lack of quantification for the merits of the decisions. In this context, spectra, especially those obtained from in-situ measurements, have become the key to addressing this challenge by providing the basic structural information of molecules thus offering feedback for each decision. Therefore, it is urgent to develop efficient methods to elucidate molecular structures on the basis of spectral information, i.e., spectrum-to-structure correlation.

Leveraging its superior ability to process big data and uncover latent patterns, deep learning (DL) has significantly advanced the spectrum-to-structure tasks. These DL-based methods can be generally categorized into two ways: database–dependent retrieval and database–independent generation. Retrieval-based approaches, including spectrum-spectrum and spectrum-structure retrieval, rely on comparing the to-be-determined spectrum with candidate spectra or molecular structures according to certain rules to find the best match. These approaches are effective in identifying chemicals within the library that has been previously established or delineated on the basis of prior knowledge, such as DeepSearch\cite{yu2025towards}, FastEI\cite{yang2023ultra} and CReSS\cite{yang2021cross}. However, these methods inevitably face severe limitations when dealing with out-of-library compounds, owing to the big gap between available experimental spectrum-structure pairs ($\sim 10^6$) and vast chemical space\cite{reymond2015chemical}. In contrast, generation-based approaches, including conditional generation and de novo generation, seek to predict molecular structures directly from spectra, bypassing the establishment and retrieval of databases. These approaches have shown great promise for predicting previously unidentified chemicals\cite{hu2024accurate,litsa2023end,wu2025transformer,stravs2022msnovelist,tran2017novo,mao2023mitigating,qiao2021computationally,alberts2024leveraging}. However, the spectral signal obtained from single technique unveils only a partial view of molecular structure. As a result, the process of converting one type of spectral data into its molecular structure is inherently challenging, let alone the complexity and noise in the experimental spectrum.

Indeed, retrieval is efficient enough to determine in-library molecules, whereas generation becomes the only option for interpreting spectra of out-of-library molecules. However, up to now most of the existing methods have either retrieval or generation but not both. Such a paradigm not only makes model unable to provide appropriate solutions as prior knowledge and databases change, but also ignores the synergy between retrieval-based and generation-based spectrum-to-structure tasks, while this synergy could further improve the performance of spectral annotation. As a result, it is ideal to develop a unified framework capable of simultaneous retrieval and generation, providing dynamic solutions on the basis of available knowledge and databases while ensuring minimal overhead in parameter volume and training costs.

In this study, we propose a DL-based \underline{\textbf{vib}}rational spectrum-\underline{\textbf{to}}-\underline{\textbf{mol}}ecular structure framework (Vib2Mol) to flexibly address a variety of spectral annotation tasks according to the available prior knowledge. Built upon an encoder-decoder Transformer architecture, Vib2Mol utilizes a multiphase training consisting of alignment and generation phases. By employing multi-task learning at each phase, the framework effectively unifies diverse spectrum-to-structure tasks into one comprehensive model. In addition, we implement coarse-to-fine retrieval and generate-then-rerank strategies to further enhance spectrum-structure retrieval and de novo generation, respectively. For a better evaluation, we compiled theoretical and experimental benchmarks, drawing upon both public datasets and our own Density functional theory (DFT) calculations. Overall, Vib2Mol not only achieves state-of-the-art performance on all benchmarks but also exhibits an overwhelming superiority when compared to mainstream methods. It further enhances the accuracy in interpreting spectra of reaction products and peptide sequencing as more knowledge of target molecules is introduced. This advancement demonstrates significant potential for in-situ intelligent analysis of dynamic chemical transformations and biological processes.

\section{Results}
\subsection{Workflow of Vib2Mol}

\begin{figure}[H]
\centering
\includegraphics[width=\textwidth]{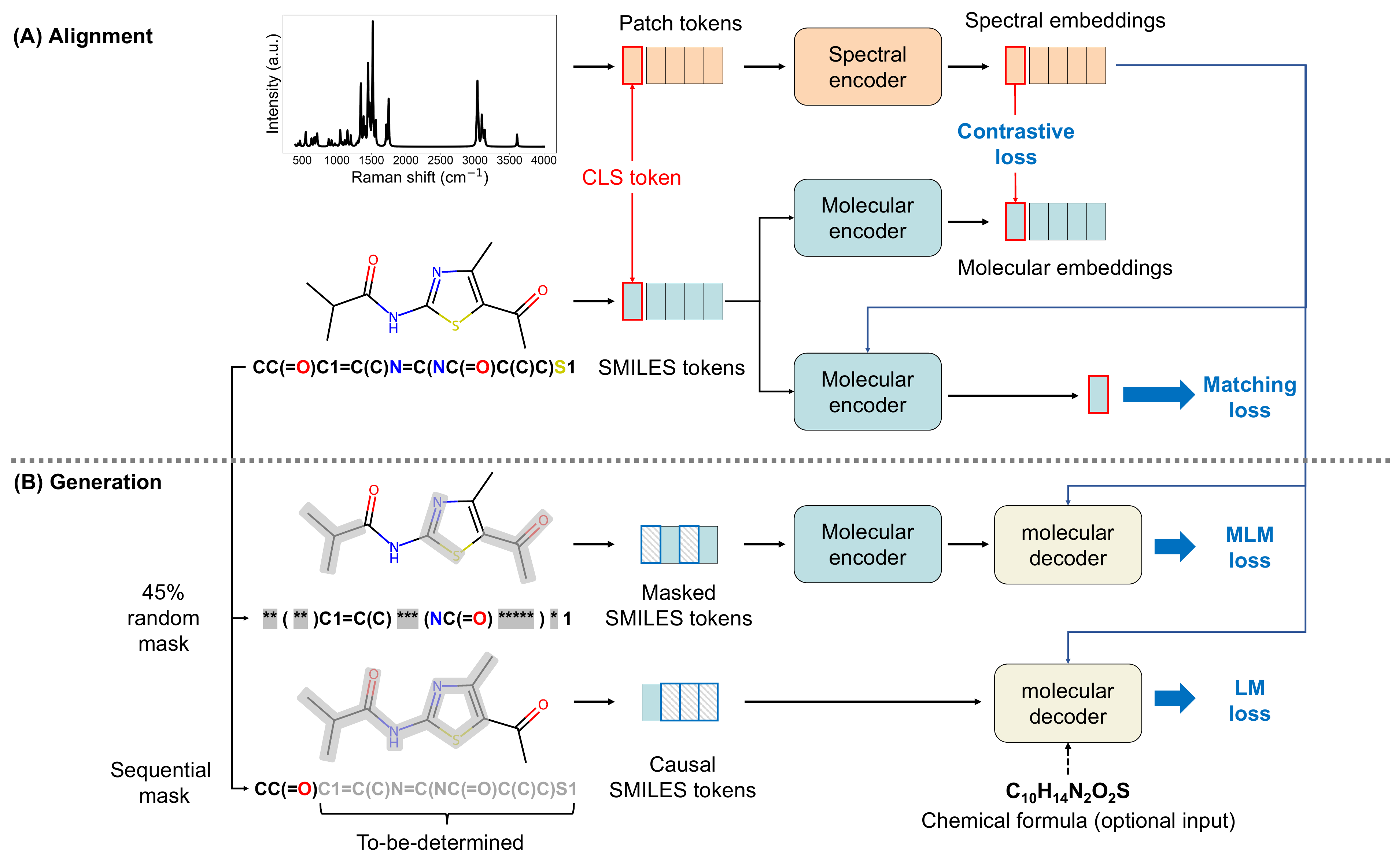}
\caption{The framework of Vib2Mol for pretraining. (A) The alignment phase: spectra and molecular structures are represented as patch tokens and SMILES tokens, respectively. After processed by their encoders, spectral and molecular information are aligned by CL. Subsequently, hard negative samples are selected and employed to guide model in learning the subtle distinctions between these highly similar spectra or molecule samples. (B) The generation phase: for conditional generation, molecules are randomly masked 45\% and encoded by the same molecular encoder used for spectrum-structure alignment. The molecular decoder fuses spectral information with molecular features and predicts masked tokens. For de novo generation, molecule is sequentially masked and directed input into the same molecular decoder as conditional generation without the prior encoding. Then, the decoder predicts the next token on the basis of previous information, spectral features and chemical formulae (if given).}
\label{fig:figure1}
\end{figure}

The workflow of Vib2Mol during training, including alignment and generation phases, is illustrated in Figure \ref{fig:figure1}. Vib2Mol adopts \underline{m}ulti\underline{p}hase \underline{t}raining (MPT) and employs multi-task learning during each phase. The first alignment (Figure 1A) aims to bring the spectral and structural features of the same molecule as close as possible while separating the features of different molecules simultaneously. Spectra and molecular structures are represented as patch tokens and SMILES tokens, and then encoded into spectral and molecular embeddings by encoders, respectively. These two embeddings are effectively aligned through contrastive learning (CL), enabling cross-modal spectrum-structure retrieval. To further enhance retrieval performance, we deliberately selected hard negative samples—highly similar molecule-spectrum pairs—from each training batch. A matching loss was then utilized to guide the model in learning the subtle distinctions inherent in these challenging samples.

Figure 1B depicts workflow of the second phase, including conditional generation and de novo generation of molecular structures. Conditional generation, i.e., predicting the masked molecular structure on the basis of the spectrum, draws on masked language modeling (MLM). Briefly, SMILES tokens, representing the molecular structure, are randomly masked by 45\% and then processed by molecular encoders to generate molecular features. The spectral and molecular encoders were initialized by cloning the parameters from the alignment phase depicted in Figure 1A. These parameters were then subsequently fine-tuned specifically for the generation task to optimize performance. Then molecular decoders fused information from both masked molecular embeddings and spectral features, and predicted the to-be-determined tokens using cross-attention. Differently, de novo generation draws on language modeling (LM). SMILES tokens are sequentially masked from left to right and directly input into the molecular decoders sharing parameters with MLM. Guided by spectral features and previously generated SMILES sequences, the decoders can predict the next SMILES token from left to right until the entire sequence is complete. Note that the chemical formula is an optional input, as vibrational spectroscopies like Raman and IR are inherently limited in determining molecular formulas. While including the formula unequivocally enhances performance, Vib2Mol remains capable of achieving commendable results in its absence (Section 2.4).

Figures 2A to 2E illustrate the workflow of Vib2Mol during inference. (1) For spectrum-spectrum retrieval (Figure 2A), instead of directly comparing spectral similarity by metrics such as Pearson correlation coefficient, the to-be-determined spectrum is encoded into an embedding vector, and the cosine similarity is calculated between this vector and the known spectral embedding vector in the database. (2) Spectrum-structure retrieval (Figure 2B) leverages coarse-to-fine retrieval strategy. Candidate molecules are identified by calculating the similarity between the embedding vector of the query spectrum and those of known molecules within the database (coarse retrieval). Subsequently, the matching module acts as a reranker to identify the most accurate structure from the candidate pool (fine-grained reranking). (3) For conditional generation (Figure 2C), Vib2Mol adopts the encoder-decoder architecture. Both the spectrum and partially masked molecular structure are encoded and then fused through the molecular decoder to generate the SMILES of the masked part. (4) For de novo generation (Figure 2D), Vib2Mol follows a generate-then-rerank strategy. In the generation step, the molecular decoder sequentially predicts each character of the SMILES string on the basis of the encoded spectral features. During token generation, beam search (see Methods for details) is employed to maintain output diversity and enhance generative performance. Subsequently, consistent with the pipeline of spectrum-structure retrieval, the matching module acts as a reranker to identify the most accurate structure from the candidate pool. Figure 2E illustrated the implementation of reranker. Candidate molecules were filtered with their chemical formula, and then a pretrained molecular encoder functions as a matching module, generating scores from a comprehensive evaluation of the query spectrum and candidate molecule features. Only candidates exhibiting high matching scores are subsequently selected as the final results. Even in the absence of chemical formula, the reranking module can still effectively sort candidates relying solely on its model-based scoring.

TIt is noteworthy that extensive parameter sharing and feature reuse are employed during both training and inference. Consequently, Vib2Mol not only operates with less active parameters (see Methods for details) but also effectively addresses these four spectrum-to-structure tasks. Among these, spectrum-structure retrieval is more effective than spectrum-spectrum retrieval because it makes better use of molecular databases\cite{lu2024deep}. De novo generation offers more flexibility than conditional generation as it does not require a predefined molecular scaffold to predict side-chain structures. Therefore, for the spectrum-to-structure problem, spectrum-structure retrieval and de novo generation constitute the most versatile solutions, and will be the primary focus of following discussion.

\begin{figure}[H]
\centering
\includegraphics[width=\textwidth]{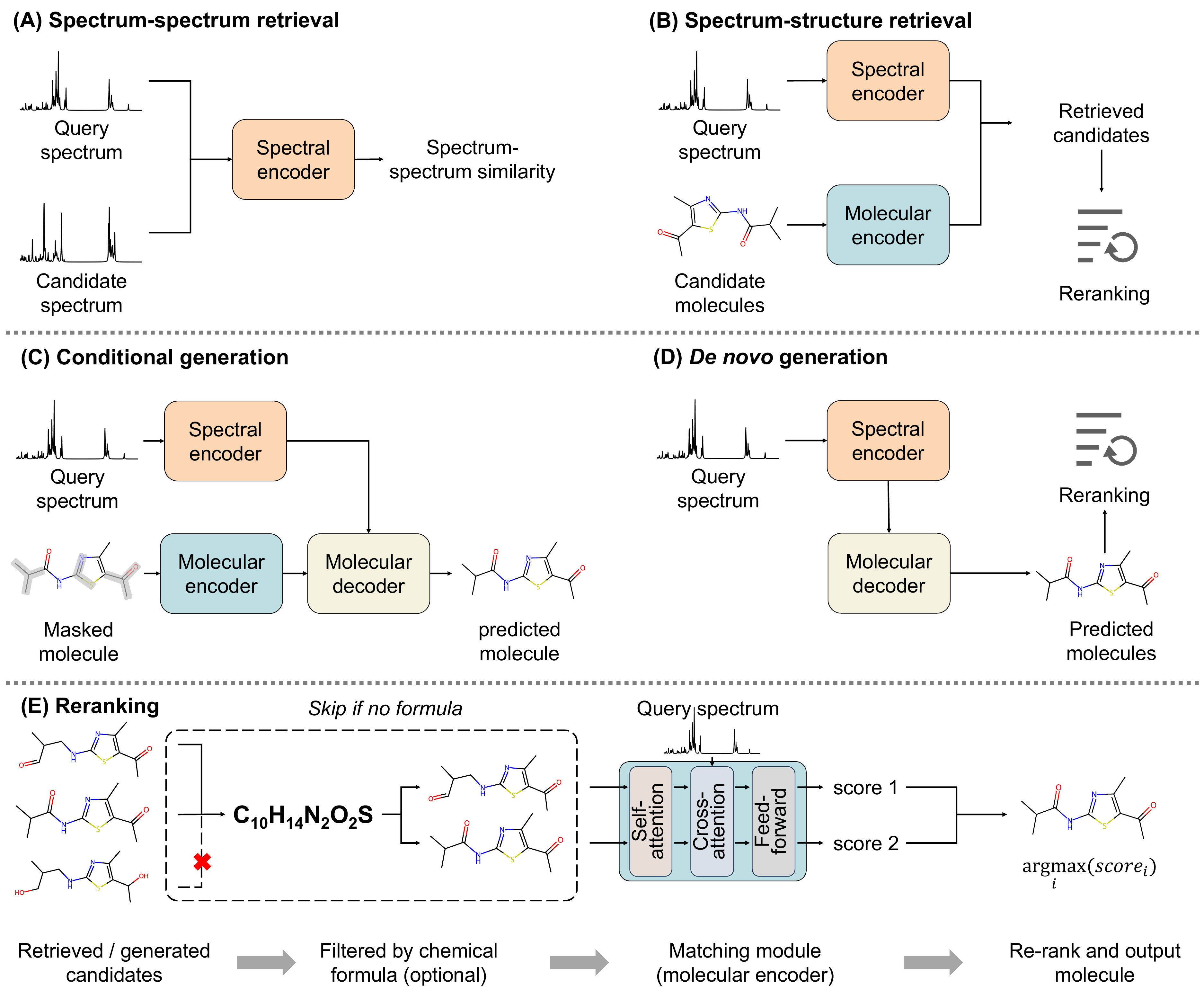}
\caption{The workflow of Vib2Mol for addressing different spectrum-to-structure tasks: (A) spectrum-spectrum retrieval, where only the spectral encoder is used to calculate the similarity between spectral pairs; (B) spectrum-structure retrieval, where spectra and molecules are encoded by their respective encoders to determine spectrum-structure similarity; (C) conditional generation, and (D) de novo generation, both following workflows during the stage of pretraining. (E) re-ranking module for refining retrieval and generation results. It initially filters candidates by chemical formula (if available), then uses a pre-trained molecular encoder to score them against the query spectrum. High-scoring candidates are finally selected as output.}
\label{fig:figure2}
\end{figure}
\newpage

\subsection{State-of-the-art performance of Vib2Mol}
To evaluate the performance of Vib2Mol on different spectrum-to-structure tasks and fairly compare it with current advanced models, five benchmarks are established. The benchmarks, which are described in detail in Methods, consist of three theoretical datasets (QM9S\cite{zou2023deep}, VB-Mols, and VB-GEOM) and two experimental ones (SDBS\cite{saito2011development} and NIST-IR\cite{linstorm1998nist}). Cumulatively, these benchmarks provide a total of 662,668 theoretical and 17,774 experimental spectra. There benchmarks enable a fair and comprehensive comparison with existing state-of-the-art methods, which can be broadly classified into three categories (see Methods for details): retrieval model (vibraCLIP\cite{rocabert2025multi}), generative models (PBSA\cite{wu2025transformer}, IR2Mol\cite{alberts2025setting}), and comprehensive models capable of both retrieval and generation tasks (SMEN21, Vib2Mol). The hyperparameters of these models are recorded in Table S1-S2. All benchmarking metrics are illustrated in Figure 2 and meticulously documented in Table S3-S5. Unless otherwise noted, Top-1 Recall (Recall@1) serves as the default metric throughout this section.

As depicted in Figure 3A, Vib2Mol demonstrated remarkable spectrum-structure retrieval performance on both the QM9S (98.11\% for Raman, 96.63\% for IR) and VB-Mols (94.66\% for Raman, 93.38\% for IR) benchmarks, outperforming vibraCLIP and performing on par with SMEN on QM9S (97.89\% for Raman, 97.04\% for IR) and VB-Mols (95.43\% for Raman, 94.02\% for IR). SMEN excels on theoretical benchmarks because it incorporates the precise molecular conformation as an extra input rather than simply using SMILES or 2D molecular graphs, which boosts its molecular representation and spectrum-retrieval performance. Despite this difference, Vib2Mol's performance is still quite comparable.

The spectrum-spectrum retrieval performance was further evaluated using VB-GEOM, a subset derived from the VB-Mols test set. This benchmark consists of pairs of stable conformers for each molecule, and each individual conformer is linked to a specific IR/Raman spectrum. The task is how to accurately identify the specific reference matching the query spectrum while distinguishing the query spectrum from structurally or spectrally similar incorrect matches. Vib2Mol demonstrated superior performance (77.54\% for Raman, 75.33\% for IR) among all models again, outperforming the second-place vibraCLIP (75.22\% for Raman, 70.16\% for IR) by a significant margin (Table S5).

When it comes to experimental benchmarks, all models were pretrained on VB-Mols and then fine-tuned with the corresponding training set, and finally evaluated with the test set, given the inherent data scarcity of these experimental datasets. Vib2Mol demonstrated a substantially lead in these benchmarks. Our model achieved the Recall@1 of 83.54\%, 86.17\%, and 90.43\% on NIST-IR, SDBS-IR, and SDBS-Raman, respectively, which surpassing the metrics of vibraCLIP. Notably, SMEN was excluded since it could not be fine-tuned without molecular conformation data for target spectra in experimental benchmarks.

Figure 3B illustrates the de novo generation performance of various models across all benchmarks. Vib2Mol consistently achieved state-of-the-art performance. For the QM9S and VB-Mols benchmarks, Vib2Mol outperformed the second-best model by at least 10\%. The distinguished performance of Vib2Mol can be attributed to not only multi-task learning and the multiphase training but also the generate-then-rerank strategy during inference.

However, we observed a pronounced performance gap between theoretical and experimental spectral test sets, owing to the following three factors: 1. Critical role of data quality and consistency. Different from the pristine mapping between structure and spectrum in the theoretical spectral dataset, experimental spectra (such as SDBS and NIST) are crowd-sourced and inherently heterogeneous, considering the variance in experimental conditions (e.g., energy resolution, quantum efficiency of CCD, temperature and solvent) can obscure the underlying spectrum-structure correlation; 2. Sparsity and incompleteness of experimental spectra. Experimental spectra frequently fail to provide full-range spectral coverage, unlike their theoretical counterparts; 3. Data scarcity in the experimental domain. For instance, sample sizes are several orders of magnitude smaller than theoretical datasets, which limits the ability of model to generalize complex patterns from theoretical simulations to the noisy, low-resource realities of experimental measurement, potentially leading to overfitting.

\subsection{Unified multimodal modeling and large-scale retrieval}

The impact of multimodal spectral information on the performance of Vib2Mol was further investigated. Rather than maintaining separate models for Raman-only, IR-only, or joint inputs, we developed a unified, multimodal model capable of intelligently processing any of these scenarios. This was achieved by employing \underline{m}asked \underline{m}odality \underline{m}odeling (MMM) during the training of \underline{m}ulti\underline{m}odal variant of Vib2Mol (Vib2Mol-MM). Specifically, in 50\% of the training steps, Vib2Mol-MM was provided with joint IR and Raman signals. For the remaining half, only one modality was randomly selected as input (see Methods for implementation details). This strategy not only enables the model to handle diverse input combinations seamlessly, but also reduces the cost of model storage and maintenance by shifting the spatial complexity from $O(2^M-1)$ to constant $O(1)$ where M denotes the number of spectral modalities.

Figures 3C and 3D compare the retrieval and generation performance of Vib2Mol-MM with and without MMM. A trend consistent with related work\cite{hu2025deep} emerged across all datasets: Raman-based models outperformed IR-based spectra, and multimodal approach (IR+Raman) substantially outperformed unimodal models across all evaluated benchmarks. Obviously, the complementary molecular structural information provided by these two vibrational spectra offers the model richer clues and a more integrated framework for constructing molecular structures or spectra. In addition, the impact of MMM on performance is significant, particularly for Raman-only and IR-only tasks on experimental data. Without MMM, a Raman-based model is restricted to learning spectrum-structure correlations solely from Raman data. In contrast, the introduction of MMM enables Vib2Mol-MM to learn a generalized representation of these correlations by leveraging the underlying physical information from both Raman and IR modalities, thereby enhancing the model's robustness on single-modality tasks.

A large-scale retrieval analysis was further conducted to evaluate model performance in real-world scenarios. Specifically, we created a candidate pool of 80,368,968 unique molecules from PubChem, filtering for those composed of C, H, N, O, F, S, Cl, P, and Br with a heavy atom count $\le$ 35. Retrieval performance was assessed by using experimental Raman/IR from the SDBS test set as queries against a merged library of PubChem and SDBS. The performance of Vib2Mol-MM was systematically evaluated by scaling the number of negative molecular samples from thousands to millions. As expected, the introduction of millions of hard negatives increased the difficulty for retrieval (Table S8). A critical inflection point was observed between ${10}^5$ and ${10}^6$ candidates, where Recall@1 decreased from 83.69\% to 68.44\%; subsequent orders of magnitude led to further substantial declines. Nevertheless, achieving a 30.60\% Top-1 Recall against 80 million candidates is remarkable, especially considering the random-chance probability of approximately $1.24\ \times\ {10}^{-8}$. This underscores the robust discriminative power of Vib2Mol within an exceptionally dense chemical space.

\begin{figure}[H]
\centering
\includegraphics[width=\textwidth]{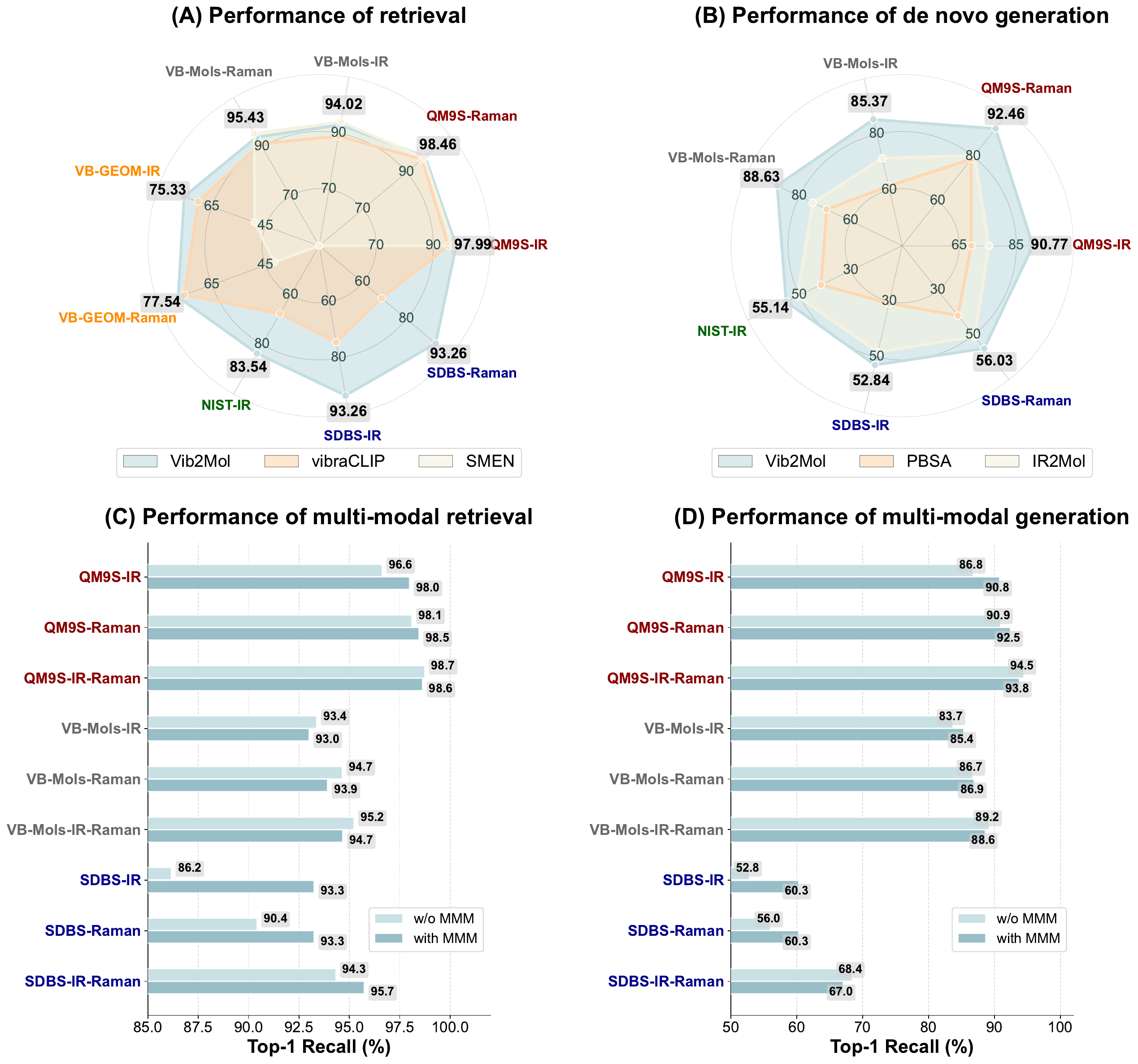}
\caption{Performance evaluation of advanced deep learning models. (A) and (B) present a performance comparison of various models on spectrum-to-structure retrieval and de novo molecular generation, respectively. These evaluations were conducted on both theoretical (VB, QM9S) and experimental (NIST, SDBS) benchmarks. The impact of multi-modal spectral input on performance of Vib2Mol is further detailed in (C) for retrieval and (D) for generation.}
\label{fig:figure3}
\end{figure}

\subsection{Module optimization and synergistic integration in Vib2Mol}

Vib2Mol consists of four modules: the retrieval (CL) module, matching (re-ranking) module, conditional generation (MLM) module, and de novo generation (LM) module. We explored the synergistic interaction among these modules across different tasks using a series of ablation experiments on the VB-Mols-Raman dataset.

For retrieval task, CL-only model achieves a Recall@1 of 88.46\% at the beginning (Figure 4A). The subsequent integration of the matching loss alone elevated retrieval performance to 89.57\%. This demonstrates that when joint losses are optimized, the matching module effectively serves as an auxiliary factor, significantly enhancing the primary retrieval task. The incorporation of chemical formulae for filtering inaccurately retrieved candidates further enhanced performance to 91.20\%. Crucially, the final re-ranking step yielded the most substantial gain in Recall@1, achieving 94.66\% (with chemical formula) and 93.20\% (without). The consistent improvement underscores the efficacy of this comprehensive strategy.

For de novo generation (Figure 4B), data augmentation for SMILES strings emerged as the most significant strategy for improving metrics, elevating Recall@1 from 60.02\% to 76.08\%. This outcome is consistent with previous research20, largely because data augmentation prevents the model from overfitting to the syntactic patterns of standard SMILES, instead guiding it to learn the intrinsic correlation between spectral data and molecular structures. Following this, the introduction of SPT and MLM loss each contributed a modest improvement to the generation metrics. For more details on the ablation study related to MLM, please refer to the Supplementary Information. Crucially, the inclusion of chemical formulae then propelled the overall performance to a new level, rising from 77.49\% to 81.93\%. This substantial gain can be attributed to the chemical formula constraints, which effectively assist the model in determining elemental composition and overall unsaturation, thereby reducing uncertainty during generation. Furthermore, employing beam search enhanced the diversity of generated outputs, successfully preventing greedy decoding from converging on local optima. When a chemical formula was supplied, a chemical formula-based filter rigorously guaranteed the validity of the generated results through rule-based enforcement. Ultimately, the re-ranking module provided an additional boost to the overall Recall@1 of 86.74\% (with chemical formula) and 82.59\% (without).

\begin{figure}[H]
\centering
\includegraphics[width=\textwidth]{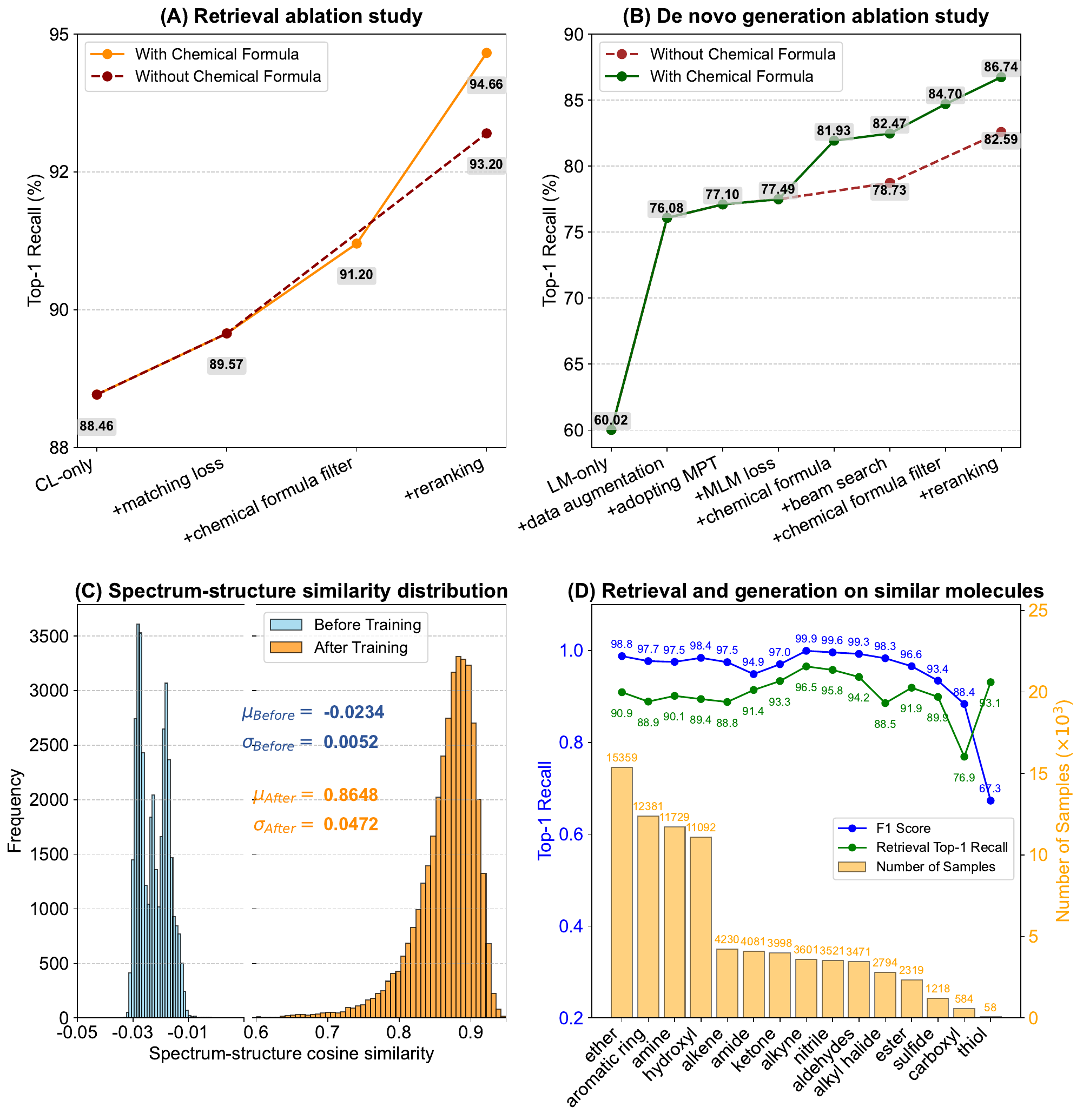}
\caption{Ablation studies of Vib2Mol on the VB-Mols-Raman dataset and visualization and statistical analysis of Vib2Mol representation learning. The performance contributions of different modules and hyperparameters are systematically assessed for three tasks: (A) retrieval and (B) de novo generation. (C) Alignment of spectral and structural embeddings, which illustrates the distribution of cosine similarities between spectrum and structure embeddings of the same molecule before and after training. (D) Performance on similar molecules categorized by functional groups, including retrieval and de novo generation tasks.}
\end{figure}

\subsection{Spectrum-structure correlation captured by Vib2Mol}

From the macro perspective, the effectiveness of Vib2Mol was explored on the test set of VB-Mols-Raman. Figure 4C visualizes the distribution of cosine similarities between spectrum-structure embeddings of individual molecule before and after training by Vib2Mol. The similarity distribution before training was concentrated around -0.023, and significantly increased to 0.86 after training, demonstrating the effective alignment between spectral and structural representations of each molecule upon training with Vib2Mol. To further investigate if our model truly understanding of the spectrum-structure relationship, we analyze its performance on failure cases. Figure S1 illustrates the Tanimoto similarity distributions between incorrectly generated or retrieved molecules and their targets. Among the incorrectly generated results, approximately 56\% of the molecules had a Tanimoto similarity greater than 0.5 to their target molecule (0.5 is a common threshold for determining molecular similarity\cite{alberts2024leveraging, hu2025deep}), with the overall distribution having an expectation of 0.53. Among the incorrectly retrieved molecules, about 47\% had a Tanimoto similarity greater than 0.5, and the expectation for this distribution is 0.47. These results suggest that even when the outputs were incorrect, they still maintained a degree of structural similarity to their intended targets, such as misplacing a methyl group or an oxygen atom.

From the micro perspective, we explored the retrieval and generation performance of Vib2Mol on similar molecules by categorizing the test set into 15 common functional groups. Figure S2 presents the t-SNE visualization of the distribution of spectral and structural embeddings within these functional group subsets. In the majority of instances, the spectral and structural embeddings of the same molecule exhibited overlap, affirming their successful alignment. Following this, the model's performance in spectrum-structure retrieval and de novo generation was assessed within each subclass (Figure 4D). For spectrum-structure retrieval, Vib2Mol achieved a weighted average Recall@1 of 90.75\% across all functional groups. In de novo generation, Vib2Mol demonstrated a Recall@1 of 86.74\% on the entire test set. Notably, when considering only the functional group accuracy of the generated molecules, the weighted average F1-score reached an impressive 97.87\%. These results collectively underscore the robust capability of Vib2Mol in distinguishing among similar molecules.

\subsection{Generating products for chemical reaction}
The autonomous robot laboratory is leading new paradigm shifts in fields such as chemical synthesis, catalysis and drug screening\cite{burger2020mobile}. The related advancements are underpinned by spectral information provided by various techniques. For instance, the autonomous and efficient exploration of chemical synthesis (such as combinatorial small-molecule synthesis, designing supramolecular materials, and screening photocatalysts) can be achieved with the aid of HPLC-MS and NMR\cite{dai2024autonomous}. However, applying current spectrum-to-structure methods to real conditions remains challenging. On the one hand, researchers have different levels of knowledge about different synthetic methods. As a result, it is crucial to fully utilize the available prior knowledge to help select appropriate molecular elucidation strategies. On the other hand, spectra measured in practice are often mixtures of reactants and products. It is a key issue to elucidate molecules under the interference of impurities. The solvation of these two problems by Vib2Mol were demonstrated as follows.

Taking the product prediction in substitution reaction of polycyclic aromatic hydrocarbons (PAHs) based on Raman spectroscopy as an example, there may have three situations (Figure 5A).

(1) Spectrum-structure retrieval is for the well-known reactions, i.e., predicting the specific substitution site with the known type of substituent. Due to the limited substitution sites of PAHs, it is possible to retrieve by traversing all possible substitution structures, thereby outputting the structure with the highest spectrum-structure similarity. As shown in Figure 5B, the Recall@1 of Vib2Mol for benzene, naphthalene, and anthracene reached 100.00\%, 98.25\%, and 99.57\%, respectively, indicating Vib2Mol can nearly perfectly perform spectrum-structure retrieval within the limited search space. Obviously, as prior knowledge decreases, the potential research space significantly increases, making it difficult to traverse all possible structures, and the generation is highly demanded.

(2) Conditional generation is for the partial known reactions, i.e., predicting the type of substituent with the known substitution. The weighted average Recall@1 of Vib2Mol is 98.95\% in predicting unknown substituent. Note that the accuracies here are somewhat inflated. Before performing conditional generation, it is necessary to design certain "blanks" for the model to "fill in", but Vib2Mol directly replaces the characters at the corresponding positions with "<mask>". This approach may leak the number of characters to be filled. Although we tried to change the number of characters corresponding to "<mask>", it is hardly to exhaust all possibilities. This shortcoming should to be addressed in the future.

(3) De novo generation is for the new reactions, i.e., predicting a completely unknown molecular structure, including the type of substituents and all substitution sites, simultaneously. Due to the simplicity of the structure, the Recall@1 of benzene (98.29\%) is significantly better than that of naphthalene (88.80\%) and anthracene (88.79\%). The weighted average Recall@1 of the three situations can reach 91.39\%. The exceptional metrics may primarily be attributed to Vib2Mol's robust alignment of spectrum-structure, particularly in the constrained generation space of PAHs.

To better reflect real-world conditions, we evaluated Vib2Mol in interpreting mixture spectra of products and reactants. By extracting approximately 15,000 chemical reactions listed in the second World AI4S Prize-Material Science Track\cite{sais2024}, we calculated the Raman spectra of reactants and products, and mixed them according to the yields (ignoring the differences in Raman scattering cross-sections of different molecules) to simulate the mixed spectra measured in real condition (see Supplementary Information for details). Spectrum-structure retrieval and de novo generation were used to simulate the scenarios where the expected product is either present in or absent from the database, respectively.  

Figure 5B illustrates that Vib2Mol achieved a Recall@1 of 98.11\% and 55.84\% on the unmixed spectrum test set for retrieval and generation tasks, respectively. However, the performance dropped considerably to 74.29\% and 36.81\% when tested on mixed spectra, highlighting that training on unmixed dataset is not suitable for mixture analysis for chemical reactions. Therefore, we trained a Vib2Mol-RXN for mixture spectra of products and reactants. Vib2Mol-RXN achieved not only a comparable performance on the unmixed spectral set, but also a significant improvement on the mixed spectrum test set (97.51\% for retrieval and 56.64\% for generation). These results clearly demonstrate that introducing yield information significantly enhances Vib2Mol’s ability to annotate mixed Raman spectra.

\begin{figure}[H]
\centering
\includegraphics[width=\textwidth]{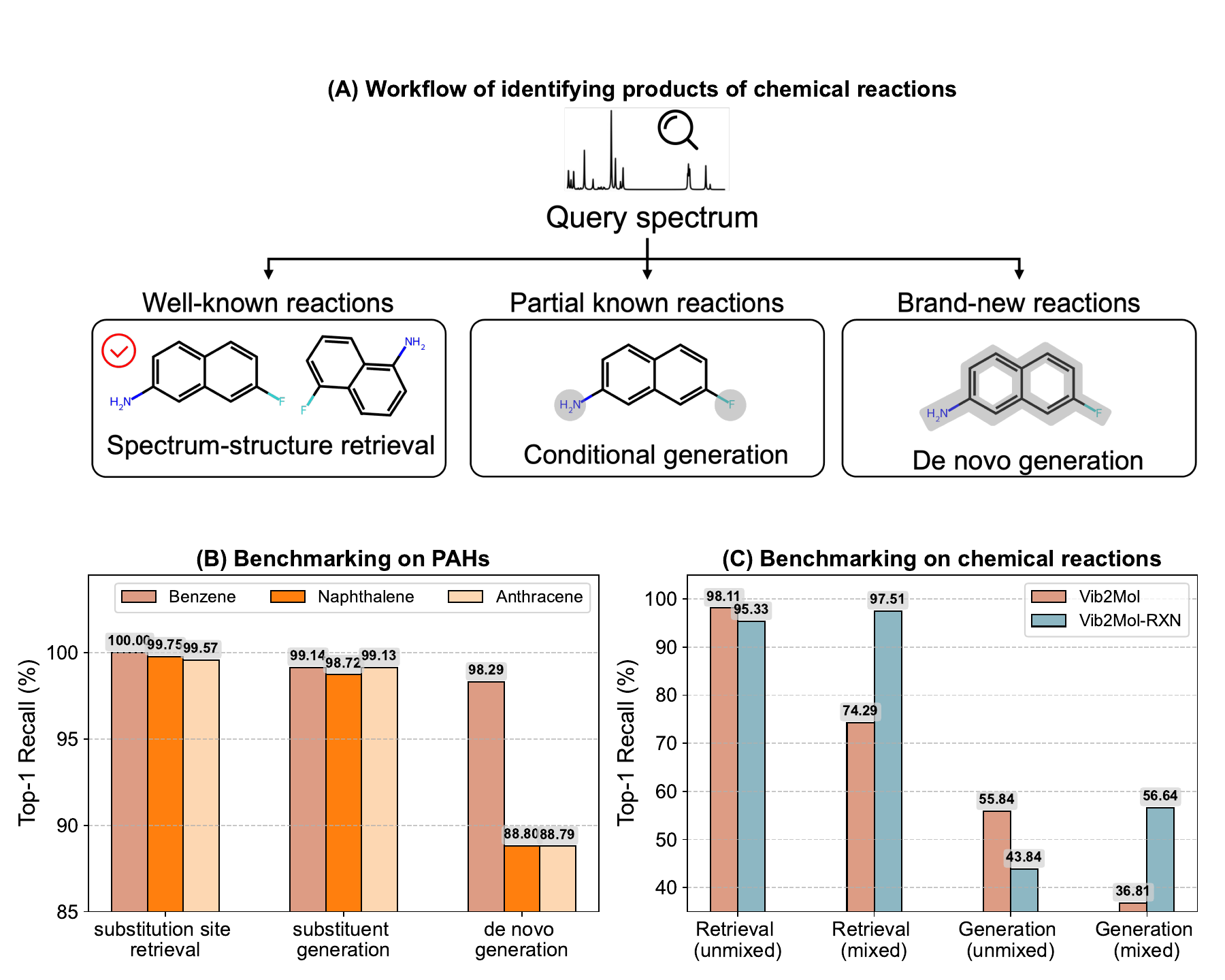}
\caption{Workflow and performance of Vib2Mol in product elucidation and mixed-spectrum analysis. (A) Three scenarios for predicting products. (B) Benchmarking on PAH substitution reactions. (C) Retrieval and de novo generation results on unmixed and mixed spectra of general chemical reactions.}
\end{figure}

\subsection{Peptide sequencing and PTMs identification}
Native proteins are composed of 20 amino acids and their post-translational modifications (PTMs). As sequences determine the structures and functions of proteins, protein sequencing and the identification of PTMs sites are key issues in reveals the functions and mechanisms of proteins in cellular function regulation, gene expression regulation, signal transduction, and the occurrence and development of diseases. To simplify the complexity of protein sequences, a bottom-up strategy is commonly adopted, which involves generating peptides of varying lengths (1-4 amino acids or longer) through chemical or enzymatic cleavage. By sequentially identifying these peptides, de novo sequencing can be achieved. However, considering the vast sequence space of polypeptides ($20^{N}_{AA}$), efficiently identifying the 20 amino acids and their combinations remains highly challenging\cite{pappas2016dynamic}. Although the unique fingerprint vibrational information in Raman spectrum is for each biomolecule (i.e., DNA, proteins)\cite{chen2018high,zhao2022label,li2022enhanced}, the complexity of Raman spectra of peptides hindered the systematic identification of polypeptides for de novo protein sequencing. Limiting the length of peptide sequences to tetrapeptides or shorter, we tried to infer peptide sequence using its Raman spectra by Vib2Mol.

The Vib2Mol pre-trained on VB-Mols was fine-tuned by peptides represented by SMILES. As shown in Table \ref{tab:table1}, the Recall@1 of the model (Vib2Mol-SMILES) for spectrum-structure retrieval is 62.97\%, when the to-be-determined peptide is in the database. Otherwise, the Recall@1 for de novo generation drops to 50.07\%. The low Recall@1 ignited us to change peptide representation from SMILES to residue sequences, considering the relatively patterned residue structure. The obtained Vib2Mol-sequence model significantly improved Recall@1 for retrieval and generation up to 70.33\% and 51.49\%, respectively. This improvement is mainly because of the drastically reduced token length by residue sequences, thereby reducing the complexity and improving the accuracy of sequence generation. 
This is the reason why current models\cite{yu2025towards,mao2023mitigating,tran2019deep} are mainly based on residue sequences.  It is not surprising that the Vib2Mol sequence performed best in elucidating dipeptides (97.37\% and 86.84\% for retrieval and generation, respectively). Since the search space increases exponentially with the number of residues, the performance of Vib2Mol-sequence gradually decreases with the increasing length of peptides. Nevertheless, even for tetrapeptides, the Recall@1 for retrieval and generation can still reach 68.03\% and 48.18\%, respectively.

As for the identification of PTMs sites, we constructed the peptide-mod dataset, which includes the most representative phosphorylation and sulfation (see Methods for details), and fine-tuned Vib2Mol-sequence by it. Depending on the level of prior knowledge in practical application, there are three cases (Table \ref{tab:table2}).

(1) Determining modification type at a specific residue site. Vib2Mol can achieve a high accuracy (74.43\%) for the three categories (sulfation, phosphorylation, and unmodified) by conditional generation. (2) Retrieval of peptides within the database. Vib2Mol can reach a Recall@1 of 65.23\% by calculating spectrum-structure similarities. (3) De novo sequencing for peptides outside the database. A Recall@1 of 33.33\% can be achieved by generation module. As a proof-of-concept, Vib2Mol demonstrates the feasibility of using theoretical Raman spectra in de novo sequencing of peptides and identifying PTMs sites. This advancement holds significant promise for applications in biomedicine, immunology, and drug development. We anticipate a synergistic integration with experimental data will further enhance its utility and uncover new insights.

To rigorously evaluate the generalization capability of the model, we benchmarked Vib2Mol against a home-made experimental Raman spectral dataset of peptides, comprising 20 amino acids, 14 dipeptides, 13 tripeptides, and 2 tetrapeptides (Figure S7). A 5-fold cross-validation on the 29 multi-residue peptide samples (excluding monomeric amino acids) was conducted, with the performance metrics summarized in Table \ref{tab:table3}.

In the retrieval task, Vib2Mol achieved an average Recall@1 of 72.00\% and Recall@3 of 86.66\%, demonstrating its proficiency in mapping noisy experimental signals to correct structural identities. Notably, retrieval Recall scaled positively with peptide length. This is primarily due to the distribution of our dataset within the latent space. As quantified in Table S9, dipeptides exhibit higher average cosine similarity. This indicates that shorter sequences reside in a “crowded” region of the representation space, where a higher density of neighboring negative samples leads to retrieval ambiguity. Conversely, longer peptides act as “informative outliers” in sparser regions, allowing for more precise identification despite their inherent spectral complexity.

In the de novo generation, the Avg. Recall@1 of13.34\% highlights the significant gap between theoretically simulated dataset and experimental one, let alone the extreme scarcity of labeled experimental Raman spectra. Further, a declined performance is for de novo generation as the peptide chain elongated. This inverse correlation is anticipated, as the chemical search space expands exponentially with each additional amino acid residue, substantially increasing the complexity of sequence reconstruction. While de novo generation on complex experimental samples remains a frontier challenge, these results provide empirical evidence of the Vib2Mol’s retrieval capabilities within real-world biological scenarios.

\begin{table}[htbp]
  \centering
  \caption{Effect of representation and length of peptide on performance.}
    \begin{tabular}{p{10em}cccc}
    \toprule
    \multicolumn{1}{c}{} & \multicolumn{2}{c}{Peptide Retrieval} & \multicolumn{2}{c}{De novo generation} \\
    \midrule
    \multicolumn{1}{c}{} & \multicolumn{1}{c}{Recall@1} & \multicolumn{1}{c}{Recall@3} & \multicolumn{1}{c}{Recall@1} & \multicolumn{1}{c}{Recall@3} \\
    \midrule
    Vib2Mol-SMILES & 62.97 & 88.79 & 50.07 & 60.66 \\
    Vib2Mol-sequence & \textbf{70.33} & \textbf{90.71} & \textbf{51.49} & \textbf{68.25} \\
    \midrule
    Dipeptide & 97.37 & 100.00   & 86.84 & 94.74 \\
    Tripeptide & 81.38 & 95.93 & 67.57 & 82.61 \\
    Tetrapeptide & 68.03 & 89.66 & 48.18 & 65.34 \\
    \bottomrule
    \end{tabular}%
  \label{tab:table1}%
\end{table}%
  
\begin{table}[htbp]
  \centering
  \caption{Performance for various PTMs types under different tasks.}
    \begin{tabular}{cccccc}
    \toprule
    \multicolumn{1}{c}{} & \multicolumn{1}{c}{site classification} & \multicolumn{2}{c}{Peptide Retrieval} & \multicolumn{2}{c}{De novo generation} \\
    \midrule
    \multicolumn{1}{c}{} & \multicolumn{1}{c}{Accuracy} & \multicolumn{1}{c}{Recall@1} & \multicolumn{1}{c}{Recall@3} & \multicolumn{1}{c}{Recall@1} & \multicolumn{1}{c}{Recall@3}\\
    \midrule
    Unmodified & 73.79 & 74.57 & 91.22 & 32.90 & 45.74 \\
    Phosphorylated & 76.04 & 62.12 & 84.79 & 35.67 & 47.24 \\
    Sulfated & 73.73 & 59.70 & 82.69 & 31.94 & 46.37 \\
    Averaged & 74.43 & 65.23 & 85.90 & 33.33 & 46.44 \\
    \bottomrule
    \end{tabular}%
  \label{tab:table2}%
\end{table}%

% Table generated by Excel2LaTeX from sheet 'Sheet1'
\begin{table}[htbp]
  \centering
  \caption{Benchmarking Vib2Mol on experimental peptide Raman spectra}
    \begin{tabular}{p{5em}p{5em}p{5em}p{5em}p{5em}}
    \toprule
    \multicolumn{1}{c}{} & \multicolumn{2}{p{10em}}{Peptide Retrieval} & \multicolumn{2}{p{10em}}{De novo generation} \\
    \midrule
    \multicolumn{1}{c}{} & Recall@1 & Recall@3 & Recall@1 & Recall@3 \\
    \midrule
    Dipeptide & \multicolumn{1}{c}{64.28} & \multicolumn{1}{c}{78.57} & \multicolumn{1}{c}{28.57} & \multicolumn{1}{c}{28.57} \\
    Tripeptide & \multicolumn{1}{c}{76.92} & \multicolumn{1}{c}{92.31} & \multicolumn{1}{c}{0} & \multicolumn{1}{c}{7.69} \\
    Tetrapeptide & \multicolumn{1}{c}{100} & \multicolumn{1}{c}{100} & \multicolumn{1}{c}{0} & \multicolumn{1}{c}{0} \\
    5-fold avg. & 72.00±9.56 & 86.66±12.47 & 13.34±12.47 & 16.66±18.25 \\
    \bottomrule
    \end{tabular}%
  \label{tab:table3}%
\end{table}%

\newpage

\section{Discussion}

In this study, we proposed Vib2Mol, a DL model for vibrational spectroscopy, which can effectively address multiple spectrum-to-structure tasks according to available prior knowledge. It not only achieves state-of-the-art performance in analyzing theoretical Infrared and Raman spectra, but also outperform previous models at experimental data. Such outstanding performance stems from the synergistic of retrieval and generation modules which lead to the better establishment of the correlation between spectrum and molecular structure.

Vib2Mol has shown substantial potential in chemical and biological applications, where we have further tackled several unexplored challenges with in-silico data. On the one hand, chemical reactions inevitably lead to a mixture of reactants and products, which thus results in mixed spectra posing a challenging issue for spectral annotation. We showcased the capability of Vib2Mol to interpret mixed-spectra, which achieved the recall@1 of 97.51\% and 56.64\% for retrieval and de novo generation on chemical reaction dataset with real yields, respectively. On the other hand, Vib2Mol not only achieved a Recall@1 of 39.9\% for de novo peptide sequencing, but also efficiently predicted PTMs sites of phosphorylated and sulfated modification, where the traditional mass spectrometry falls short, which enables Raman spectroscopy as a unique and promising omics method.

Vib2Mol also demonstrates the potential for in situ monitoring of dynamic chemical reactions and life processes using vibrational spectroscopy, particularly in discovery-oriented and automated experimental settings, where reaction pathways, intermediates, or products cannot be predefined a priori. In the future, to better elucidate molecular conformations in dynamic processes, a possible improvement lies in the introduction of stereochemical information. Furthermore, the current framework relies heavily on a SMILES-based molecular encoder, which restricts its applicability to non-molecular systems such as polymers and minerals. Transitioning toward more flexible molecular representations, such as graph-based or 3D-aware encoders, would broaden the application of this method. Finally, there is significant potential in developing bidirectional generative modules that enable seamless transitions between spectrum-to-structure and structure-to-spectrum predictions, creating a truly reversible mapping for molecular design and analysis.

\section{Methods}

\subsection{Reference data}

We have developed a \underline{\textbf{vi}}brational spectrum-to-structure \underline{\textbf{bench}}mark (ViBench, VB), which consists of two parts: VB-Mols and VB-geometry (VB-GEOM), which can be employed as a lead board for fairly comparing current advanced deep learning models. Additionally, five more benchmarks are established using publicly available datasets to further enhance the reliability of our evaluation. In total, these benchmarks comprise 662,668 theoretical and 17,774 experimental spectra.

From the perspective of data source, these benchmarks are categorized into to theoretical and experimental benchmarks. Theoretical benchmarks include QM9S, VB-Mols, and VB-GEOM. These datasets provide simulated infrared and Raman spectra along with optimized molecular conformations. Among them, QM9S serves as the most basic and widely used theoretical benchmark. It consists of small organic molecules (with fewer than 9 heavy atoms) and their corresponding IR and Raman spectra. The molecules in this dataset are composed exclusively of C, H, O, N, and F elements. Extensionally, ViBench incorporates additional Cl, Br, P, and Si elements, and increases the maximum number of heavy atoms to 45. Experimental benchmarks comprise SDBS (Spectral Database for Organic Compounds) and NIST-IR (United States National Institute for Science and Technology). It is important to note that only SDBS offers both infrared and Raman spectra simultaneously, whereas NIST-IR only provides infrared spectra. Furthermore, molecular conformation information is absent from these experimental datasets.

From the perspective of validation methodology, VB-GEOM is a unique benchmark. It comprises 6,835 molecules, each containing two distinct stable conformers, with each conformer having its corresponding IR and Raman spectra. Thus, each data entry in VB-GEOM is effectively a tuple including 7 elements: SMILES, Conformer\#1, IR of Conformer\#1, Raman of Conformer\#1, Conformer\#2, IR of Conformer\#2, Raman of Conformer\#2. Taking Raman spectroscopy as an example, we treat the spectrum of Conformer\#1 as the query and Conformer\#2's as the reference. For retrieval tasks, the model must accurately identify the reference corresponding to the input query, excluding interference from other similar spectra. For generative tasks, the model needs to generate a complete molecular structure based on the input query. The data source, GEOM, is distinct from the QM9 and ZINC15 datasets, and all evaluations are performed under zero-shot conditions.

Moreover, to demonstrate the promising potential of our model for chemical reaction product prediction and peptide sequencing, VB-PAHs, VB-RXN, VB-peptide, and VB-peptide-mod are developed. Details are listed in supplementary information.

Density functional theory (DFT) was employed to perform conformational optimization of molecules and calculated the corresponding infrared and Raman spectra in VB. Unless otherwise specified, all quantum chemical calculations were carried out using the Gaussian 09 program\cite{g09}. The geometries were optimized using the B3LYP-D3BJ functional with a 6-311+G** basis set. Frequency calculations were obtained at the same level at the optimized geometry.

Our datasets were partitioned based on unique identifiers (GDB9-ID for QM9S and a combination of GDB9-ID/ZINC-ID for VB-Mols, etc.). We employed a stratified random sampling approach based on these IDs to construct the training, validation, and test sets. Detailed statistics regarding the specific dataset divisions are provided in Table S10. To facilitate subsequent calculations, the spectral dimensions were unified to 1024, and molecular structures were all represented using SMILES.

\subsection{Related works and baseline models}

To fairly compare the performance among Vib2Mol and current methods, we surveyed spectrum-to-structure models based on vibrational spectroscopy in the community. For spectral-structure retrieval, CL is currently the most popular framework. DeepSearch\cite{yu2025towards}, CReSS\cite{yang2021cross},  SMEN\cite{kanakala2024spectra} and vibraCLIP\cite{rocabert2025multi} all employ CL to bring the spectra and corresponding structures of the same molecule closer together, achieving great spectral-structure retrieval performance for mass spectrometry, NMR, IR, and Raman respectively. We fully followed the training and inference configurations of SMEN and vibraCLIP to test their performance on the aforementioned benchmarks. It is worth noting that SMEN, vibraCLIP, and Vib2Mol represent molecules with coordinates (conformation), molecular graphs, and SMILES, respectively. Conformation offers a significant advantage in theoretical spectrum-structure retrieval because a precisely corresponding conformation provides a better molecular representation. However, accurate conformations are unavailable for experimental benchmarks, making only conformation-independent molecular representations easily transferable.

For conditional generation, MLM is currently the most popular approach. CO-BERT realized bidirectional prediction between molecular structures and vibrational spectra by MLM\cite{yang2024cross}. However, CO-BERT focus on predicting atomic coordinates by vibrational spectra and contextual structural information. Therefore, we built a similar variant based on BERT and compared it with Vib2Mol in Supplementary Information.

For de novo generation, both IR2Mol\cite{alberts2024leveraging} and the Patch-based Self-Attention (PBSA) model\cite{wu2025transformer} employ an encoder-decoder architecture and integrate molecular formula constraints. We followed the established training and inference protocols of IR2Mol and PBSA to assess their performance across the benchmarks. It is important to note that both of these baselines employ sophisticated data augmentation strategies, and PBSA additionally uses an ensemble learning approach. While we acknowledge the efficacy of these strategies, for the sake of fairness and efficiency in this study, we exclusively used data augmentation applied only to SMILES strings and did not incorporate any ensemble learning strategies in this paper. Recognizing that molecular formulae are not always easily obtainable\cite{hu2024accurate}, we developed two distinct versions of Vib2Mol—one that incorporates chemical formulae and another that operates without them. Both versions exhibited strong performance.

Given that datasets like SDBS and NIST-IR are dynamically updated, all benchmarks presented in this paper were constructed using the latest data available as of July 2025. Moreover, to ensure a fresh evaluation, all benchmarks (including QM9S) underwent a complete re-shuffling, rather than adhering to the dataset divisions employed in prior studies.

\subsection{Spectral and molecular representation}

As shown in Figure S3A, the convolutional kernels with size of 8 were first used to slice the original spectra into 128 patches. linear projection was then employed to transform each patch into a 768-dimensional vector, i.e., spectral embeddings. As shown in Figure S3B, the preprocessing of molecules is similar. the molecular structure is represented as a SMILES string and is split into several discrete characters, i.e., SMILES tokens. After looking up the codebook, all characters are mapped to 768-dimensional vectors, i.e., molecular embeddings. Subsequently, the \textless CLS\textgreater token, representing the global information of the sequence, was inserted at the beginning of both the spectral sequence and the molecular structure sequence, and positional encoding was added to both. Finally, a 6-layer Transformer encoder based on self-attention was used to update the features of each token in the sequence. It is worth noting that at this point, the spectrum and molecular structure only interact with their own features and do not communicate with each other here. 

\subsection{Alignment between spectrum and molecular structure}
To align the features of spectra and molecular structures, CL was introduced. As shown in Figure S6, spectral and molecular features were extracted from their respective encoders, then a spectrum-structure similarity matrix was obtained through the dot product. By optimizing this matrix, the spectral and molecular embeddings of the same molecule were made as close as possible (with the diagonal elements approaching 1), while the embeddings of mismatched spectrum-molecule pairs were made as distant as possible (with the off-diagonal elements approaching 0).

During the training phase, we used a symmetric cross-entropy loss\cite{radford2021learning} to calculate the similarity errors between the spectra and structures of the same molecule and updated the neural network based on this. The specific formula of the loss function is as follows:
\begin{equation}
    L_{total}=\frac{1}{2}(L_{spectrum}+L_{structure})=-\frac{1}{2B}(\sum_{i=1}^{B}{log(p_{i})}+\sum_{j=1}^{B}{log(q_{j})})
\end{equation}
Expanding this with the similarity scores, the formulation becomes:
\begin{equation}
    L_{total}=-\frac{1}{2B}(\sum_{i=1}^{B}{log(\frac{exp(s_{i,i}/\tau)}{\sum_{j=1}^{B}{exp(s_{i,j}/\tau)}})}+\sum_{j=1}^{B}{log(\frac{exp(s_{j,j}/\tau)}{\sum_{i=1}^{B}{exp(s_{i,j}/\tau)}})})
\end{equation}
where $B$ denotes the size of mini-batch, $p_i$ and $q_j$ represent the predicted probabilities of correct alignment of spectrum-to-structure, and structure-to-spectrum. These probabilities are calculated based on the similarity scores $s_{i,j}$ and scaled by a temperature hyperparameter $\tau$.

During the testing or inference phase, only the dot product of the features of the to-be-determined spectrum and the molecules in the library needs to be calculated, and the top-k results are taken as the results (Figure S4).

\subsection{Spectrum-structure matching and re-ranking}

Building on a large-scale spectrum-structure alignment achieved through contrastive learning, we found that a small number of negative pairs—those with similar spectra or molecular structures—were still difficult to distinguish. To address this, we introduced the spectrum-structure matching. This module is a binary classification task, where the model uses a linear layer to predict whether a spectrum-structure pair is positive (matched) or negative (unmatched) given their multimodal feature. In order to find more informative negatives, we adopt the hard negative mining strategy by BLIP\cite{li2022blip}, where negatives pairs with higher contrastive similarity in a batch are more likely to be selected to compute the loss. During inference, this module naturally scores the retrieved or generated molecular structures against a to-be-determined spectrum, enabling a deep learning-based re-ranking of candidate molecules.

The matching loss is formulated as a binary cross-entropy objective:
\begin{equation}
    L_{matching}=-\sum_{k\in\{pos,neg\}} y_k\log{\left({\hat{y}}_k\right)}
\end{equation}
where $y_k$ denotes the ground-truth binary label and $\hat{y}_k$ represents the predicted probability of a match.

\subsection{Spectrum-guided molecular generation}

After aligning the spectra and molecular features, we aim to generate molecular structures based on spectral information. During the training phase, we integrated two training tasks, MLM and LM. For MLM, we randomly masked 45\% of the content in the structural sequence and utilized cross-attention to enable the model to learn how to restore the masked parts of the structure based on the spectrum and contextual tokens (Figure S3C). For LM, we enforced the model to learn how to predict the next character based on the previously generated text and under the guidance of the spectrum (Figure S3D).

The loss functions for both MLM and LM are based on cross-entropy, as detailed below:
\begin{equation}
    L_{MLM} =-\frac{1}{N}\sum_{i=1}^{N}{logP(\hat{y}_i|y_{unmasked},s)}
\end{equation}

where $N$ is the total number of masked positions, $y_{unmasked}$ represents the contextual tokens around the masked ones, $\hat{y}_i$ represents masked tokens to be predicted, and $s$ is the input spectrum.
\begin{equation}
    L_{LM} =-\frac{1}{M}\sum_{i=1}^{N}{logP(\hat{z}_j|z_{prev},s)}
\end{equation}
where $M$ is the length of the SMILES sequence, $z_{prev}$ represents the previous generated SMILES, $\hat{z}_j$ represents the next to-be-predicted token, and $s$ is the input spectrum.

\subsection{Beam Search}
For de novo generation, the greedy search strategy, which only selects the token with the highest probability as the next character may fall into local optima. To enhance the diversity of the generated results, we adopted the beam search strategy, of which the implement is derived from PBSA.

Our method works as follows: at each decoding step, the model calculates the log-probabilities for all possible next tokens. Instead of picking only the most probable token, it keeps track of the top k most probable sequences, where k is the beam size. These sequences are then expanded in the next step, and the process is repeated. The scores for the candidate sequences are accumulated as the sum of their log-probabilities. This parallel exploration of multiple promising paths helps to find better solutions that a greedy approach might miss. To control the generated results, a temperature parameter ($\tau$) is applied to the log-probabilities before the top-k selection. A higher temperature value makes the probability distribution flatter, increasing the randomness and diversity of the selected tokens. Conversely, a lower temperature value sharpens the distribution, leading to a more deterministic search similar to greedy decoding. This allows us to balance between fidelity to the most likely sequence and the exploration of diverse alternatives.

This process can be described by following equations:
\begin{align*}
    P(w_t|W_{<t},X) &= \text{softmax} \left( \frac{\text{logits} (w_t|W_{<t},X)}{\tau} \right) \\
    \text{Score}(W_T) &= \sum_{t=1}^T \log P(w_t|W_{<t},X) \\
\end{align*}
where $W_T=(w_1,w_2,...,w_T)$ is a candidate sequence of length $T$, $X$ is the input sequence, $W_{<t}=(w_1,w_2,...,w_{t-1})$ represents the sequence of tokens generated up to step $t-1$, and $\text{logits}(w_t|W_{<t},X)$ are the unnormalized log-probabilities for the next token $w_t$.

\subsection{Multimodal integration and mask modality modeling}
The multimodal variant, Vib2Mol-MM, is designed to handle heterogeneous spectral data flexibly. The spectral input is structured as a tensor ${X}\in\operatorname{R}^{B\times C\times L}$, where $B$ is the batch size, $C$ denotes the number of spectral modalities (channels), $L$ represents the spectral dimension (the number of sampling points, which is 1024 in this work). We use a 1D convolution with an equal kernel size and stride to convert the input tensor into patch tokens. whose feature dimension is $D$ which is 768 in our case.

Raman and infrared spectra share a common x-axis (wavenumber), allowing them to be treated as complementary feature channels—analogous to the RGB channels in image recognition. This shared dimensionality enables a seamless transition between modalities by simply adjusting the input channel parameter, $C$, without requiring structural architectural changes. For unimodal input, $C=1$; for multimodal input, $C=2$, as the Raman ($S_{Raman}$) and infrared ($S_{IR}$) spectra are concatenated along the channel dimension. From a machine learning perspective, this approach allows the model to learn cross-modal correlations at each specific vibrational frequency. However, it is important to acknowledge that simple channel-wise concatenation is a relatively primitive approach to multimodal integration. It may not fully account for the Rule of Mutual Exclusion or the fundamental differences in selection rules—specifically, that IR transitions depend on changes in dipole moment, whereas Raman transitions depend on changes in polarizability.

To enable a unified framework capable of processing Raman-only, IR-only, or joint inputs, we propose a masked modality modeling training strategy. As detailed in Algorithm 1, we treat the two modalities as channels. During training, we randomly mask one modality with a probability $p_{mask}=0.5$, replacing the signal with a learnable embedding $E_{mask}$. This forces the model to learn robust inter-modal dependencies when both are present, while maintaining high performance when only a single modality is available.

\begin{algorithm}
\caption{Training Forward with Masked Modality Modeling}\label{algo1}
\begin{algorithmic}[1]
    \Require Raman spectra $S_{Raman}$, infrared spectra $S_{IR}$, Training flag $T$, Masking probability $p_{mask}$
    \Ensure Latent spectral representations $Z$
    
    \State $X \gets \text{Concat}(S_{Raman}, S_{IR})$
    
    \If{$T$ is True \textbf{and} $\text{Rand}(0,1) < p_{mask}$}
        \State $m \gets \text{Sample}(\{0, 1\})$ \Comment{Randomly pick a modality to mask}
        \State $X[:, m, :] \gets E_{mask}$
    \EndIf
    
    \State $E \gets \text{Conv1D}(X)$
    \State $E_{patch} \gets \text{Reshape}(E) + E_{pos}$
    \State $Z \gets \text{TransformerEncoder}(E_{patch})$
    
    \State \Return $Z$
\end{algorithmic}
\end{algorithm}

\subsection{Model complexity and computational cost}

We have summarized the model size (parameter count), computational complexity (GFLOPS), and memory requirements (VRAM) in Table S11. As illustrated, Vib2Mol achieves superior efficiency through extensive parameter sharing. For instance, in retrieval tasks, the integrated matching module and molecular encoder reduce the active parameter count relative to independent architectures. Similarly, for generation tasks, our shared decoder framework significantly lowers memory overhead while maintaining robust multi-task capabilities. Note that the VRAM requirements for training are based on the hyperparameters specified in Table S1-S2.

Furthermore, we conducted a comparative benchmark of inference times using an NVIDIA RTX 4090 GPU and an Intel Xeon Platinum 8468 CPU, as summarized in Table S12. The inference pipeline consists of three primary stages: pre-retrieval/generation, filtering, and re-ranking. While the pre-retrieval/ generation and re-ranking stages are optimized for GPU acceleration, the filtering stage is executed primarily on the CPU. Our results demonstrate that while the GPU offers superior throughput, the model remains highly performant on a CPU, with retrieval tasks consistently completing in under 100 ms.

\subsection{Raman spectra acquisition and data processing}
All the samples were placed on a glass slide wrapped with aluminum foil and gently pressed with a second glass slide to form a uniform and flat layer. Raman spectra were then obtained on Invia confocal Raman microscope (Renishaw, UK) with a 785 nm excitation laser. The laser power was set to 16.9 mW, with an integration time of 10 s per acquisition, and each spectrum was obtained by averaging three accumulations. Savitzky–Golay filter with a window size of 7 and an order of 3 was used to denoise. After the subtraction of the biofluorescence background, spectra were individually normalized with the maximum peak value.

\bmhead{Supplementary information}  
Details about reference data, extra figures and tables are available in the supplementary information.

\bmhead{Acknowledgements}
This work was supported by the National Natural Science Foundation (Grant No: 22227802, 22021001, 22474117 and 22272139) of China and the Fundamental Research Funds for the Central Universities (20720220009 and 20720250005) and Shanghai Innovation Institute.

\begin{appendices}

\section{Details about Datasets}

We have established a vibrational spectrum-to-structure benchmark (ViBench, VB). As shown in Table S4, the molecular data of VibBench consists of eight parts:

\textbf{VB-QM9}: 133,434 organic small molecules extracted from QM9\cite{ramakrishnan2014quantum}, composed of C, H, O, N, and F atoms, with the number of heavy atoms less than 10. Each molecule in this subset has only one stable conformation

\textbf{VB-ZINC15}: 50,114 drug molecules extracted from ZINC15\cite{sterling2015zinc}, involving a wider range of elements, including C, H, O, N, S, F, Cl, Br, P, and Si, with the number of heavy atoms ranging from 4 to 45. Notably, since the ZINC15 dataset contains many isomers, and VB-zinc15 only ensures the uniqueness of ZINC-IDs, 7,556 molecules in this subset have multiple stable conformations.

\textbf{VB-mols}: For convenience in pre-training and evaluation, we merged VB-qm9 and VB-zinc15, and the combined dataset is referred to as VB-mols. In other words, VB-mols is not an additional dataset but an integration of existing data.

\textbf{VB-geometry}: 6835 organic small molecules extracted from GEOM\cite{axelrod2022geom}, each with two stable conformations. We randomly used the spectrum of one conformation as the query input and the other as the reference spectrum, thus constructing a test set for evaluating the model's spectrum-to-spectrum matching performance.

\textbf{SDBS}: 2815 organic molecules extracted from Spectral Database for Organic Compounds (SDBS\cite{saito2011development}), which contains experimental Raman and infrared spectra simultaneously. This data was collected up to July 1, 2025.

\textbf{NIST-IR}: 12144 experimental infrared spectrum-molecule pairs extracted from NIST Chemistry WebBook\cite{linstorm1998nist}. This data was collected up to July 1, 2025.

\textbf{VB-PAHs}: Includes 1,268 benzene derivatives, 1,853 naphthalene derivatives, and 1,175 anthracene derivatives. The substitution sites for benzene include (1,2), (1,3), and (1,4); for naphthalene, they include (1,2), (1,5), (1,8), (2,6), and (2,7); and for anthracene, they include (1,2), (2,3), and (2,6). All derivatives contain two common substituents as detailed in Table S5.

\textbf{VB-RXN}: 15,639 unique reaction data extracted from The second World AI4S Prize-Material Science Track. Each data entry includes the yield, structures, and Raman spectra of reactant 1, reactant 2, and the product. All molecules have a maximum of 20 heavy atoms and only contain C, H, N, O, F, S, Cl, P, and Br elements.

\textbf{VB-peptide}: Includes 273 dipeptides (68.25\% of all possible dipeptides), 4,058 tripeptides, and 21,624 tetrapeptides. All peptides are generated based on the permutations and combinations of A, N, D, C, Q, E, G, H, I, L, M, F, P, S, T, Y, and V.

\textbf{VB-peptide-mod}: Includes 3,815 unmodified peptides, 3,716 phosphorylated peptides, and 5,023 sulfated peptides. All peptides are either tripeptides or tetrapeptides with at most one modification site. The specific modification sites include O-phosphorylation and O-sulfation of tyrosine, serine, and threonine, as well as two different N-phosphorylation modifications of histidine.

\section{Ablation study of MLM}

To better evaluate the performance of conditional generation, we compared two metrics: token accuracy and molecular accuracy. As shown in Figure S5A, token accuracy takes each character to be predicted as the smallest granularity and assesses the model's ability to restore the masked characters. However, the same molecule can be represented by different SMILES. Therefore, molecular accuracy does not examine the correctness of each character but is designed to evaluate whether the finally predicted molecule is correct (Figure S5B). In addition, we only masked the content between “(” and “)”, so as to ensure that all parts to be predicted are complete branch structures which have clear structural information rather than random combination of characters.

The MLM model initially achieved a small improvement when we switched from an encoder-only framework (like BERT) to an encoder-decoder architecture. By increasing the training masking ratio from 15\% to 45\%, we observed a substantial performance leap from 87.91\% to 92.46\% (Figure S6A). As Figure S6B illustrates, a model trained with a specific masking ratio tends to perform best when tested with a similar masking ratio. For instance, a model trained with 15\% masking can recover each token with an accuracy of 99.36\% if the input SMILES strings are also 15\% masked. However, if the strings are masked by 75\%, the accuracy drops considerably to 80.65\%. In practical scenarios, such as the prediction of polycyclic aromatic hydrocarbon functional groups discussed in Section 2.5 of main text, the masking rate of input strings can fluctuate significantly and is not fixed. 45\% emerged as an optimal hyperparameter because models trained with a 45\% SMILES masking rate achieving the highest average token accuracy, demonstrating a balanced performance across both short and long-range conditional generation tasks. Subsequently, data augmentation applied to SMILES strings, and adopting SPT rather than de novo training, yielded slight improvements in performance. Conversely, the introduction of the LM loss led to a noticeable decline in conditional generation metrics. This suggests that during multi-objective optimization, the LM loss became dominant, negatively impacting the performance of the conditional generation task, which is primarily driven by the MLM loss.

\section{SMILES augmentation}
We utilized SMILES randomization (also known as SMILES enumeration) as a data augmentation strategy. One molecular structure can be represented by multiple equivalent SMILES strings. For example, propanal can be denoted as “CCC=O”, “C(=O)CC”, or “C(CC)=O”. We randomly sample these equivalent representations during training. This approach encourages the model to learn structural invariance, a technique previously shown in models like IR2Mol and PBSA to significantly enhance generative performance. Detailed implementation scripts are provided here.

\begin{minipage}{\hsize}%
\lstset{frame=single,framexleftmargin=-1pt,framexrightmargin=-17pt,framesep=12pt,linewidth=0.98\textwidth,language=pascal}% Set your language (you can change the language for each code-block optionally)
%%% Start your code-block
\begin{lstlisting}[label={list:smiles_aug}]
def smiles_augment(smiles, augmentation_ratio=0.5):
    if random.random() < augmentation_ratio:
        mol = Chem.MolFromSmiles(smiles)
        if mol:
            # Generate randomized Kekule SMILES
            augments = Chem.MolToSmiles(mol, 
                                      isomericSmiles=False, 
                                      canonical=False, 
                                      kekuleSmiles=True, 
                                      doRandom=True)
            return augments
    return smiles
\end{lstlisting}
\end{minipage}

\section{Figures}
\renewcommand{\thefigure}{S\arabic{figure}}
\renewcommand{\thetable}{S\arabic{table}}

\begin{figure}[H]
\centering
\includegraphics[width=\textwidth]{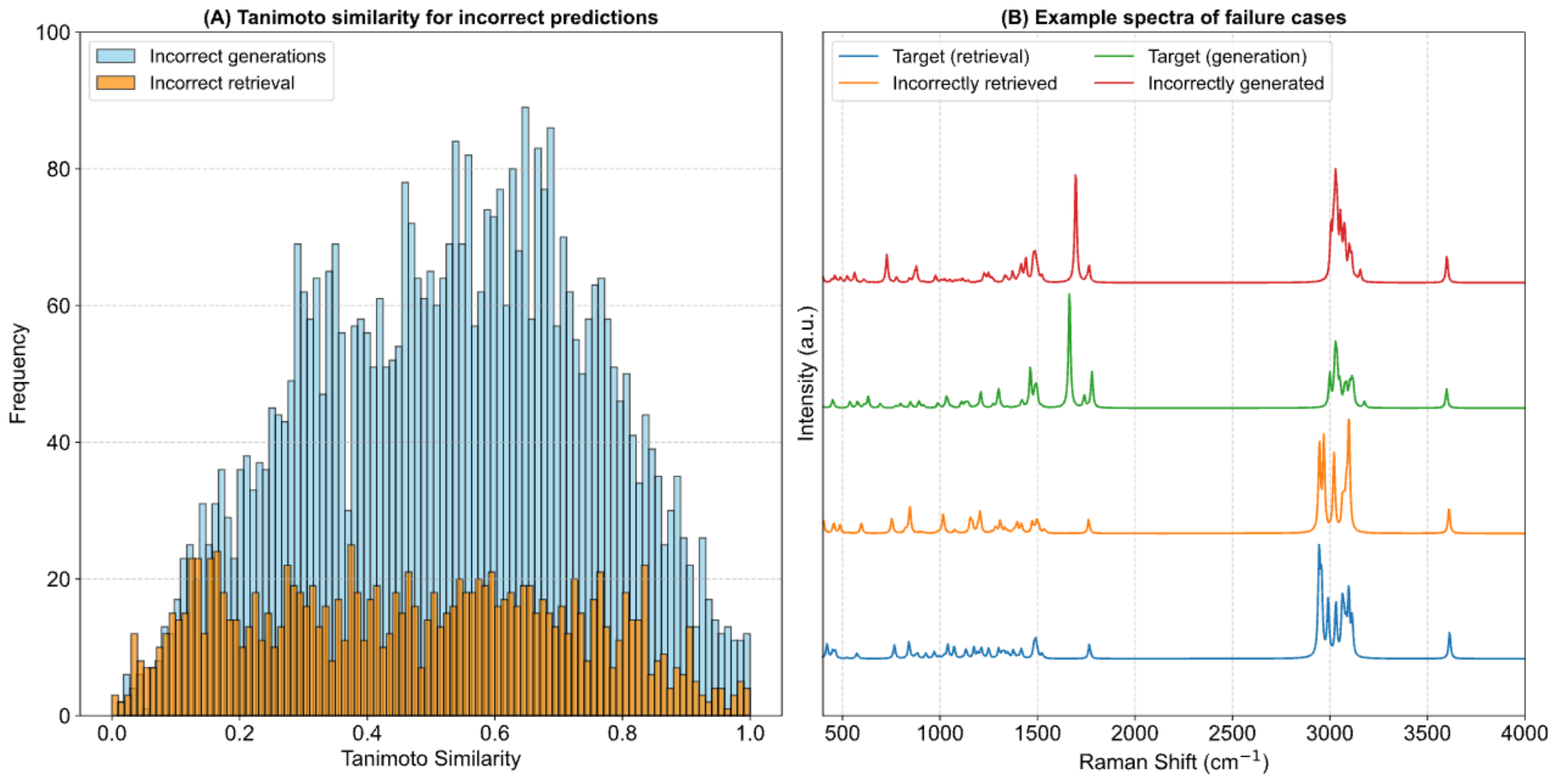}
\caption{(A) Tanimoto similarity distributions for incorrect predictions. (B) Example spectra and molecular structures of failure cases.}
\end{figure}

\begin{figure}[H]
\centering
\includegraphics[width=\textwidth]{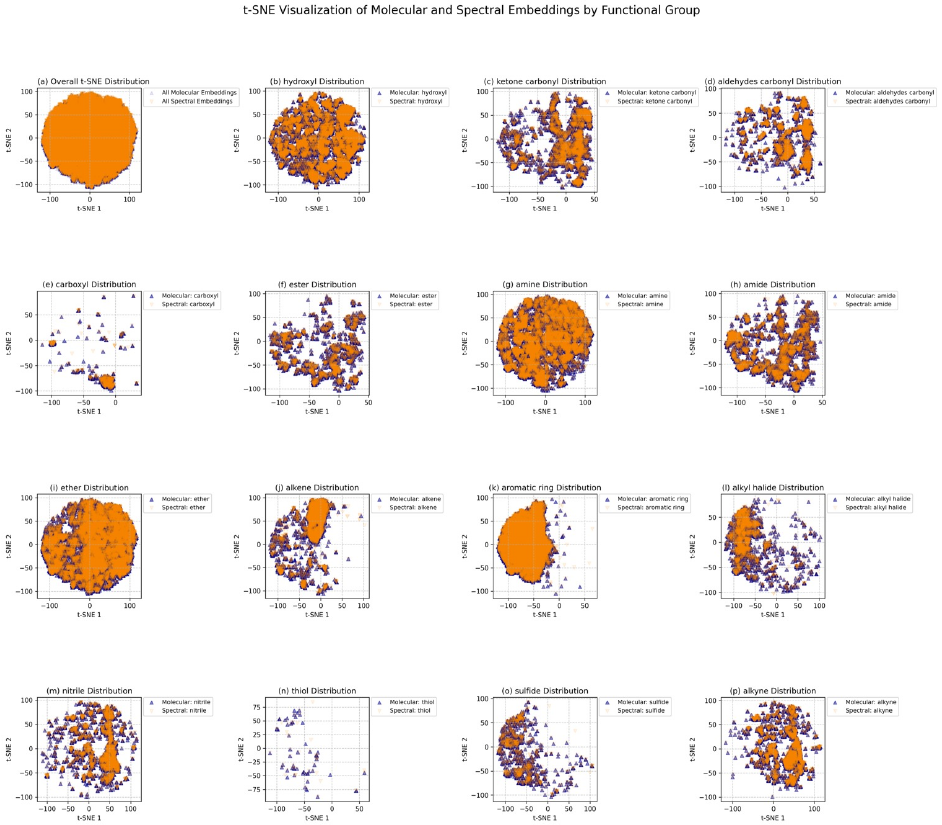}
\caption{t-SNE visualization of spectral and molecular embeddings. (A) shows the overall embeddings, while figures (B-P) show the embeddings for each functional group.}
\end{figure}

\begin{figure}[H]
\centering
\includegraphics[width=\textwidth]{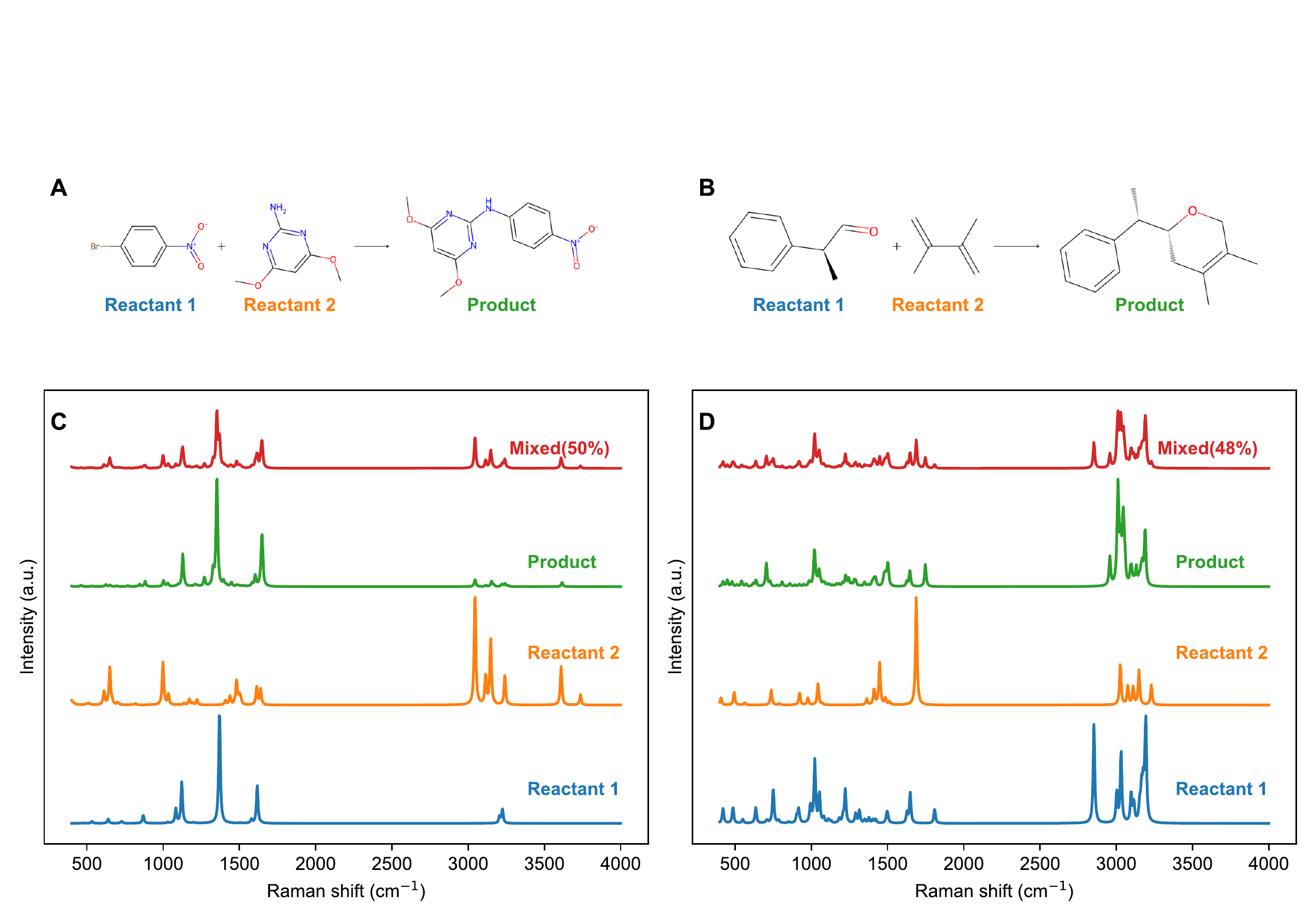}
\caption{Architectures for (A) spectral encoding, (B) molecular encoding, (C) masked language modeling and (D) language modeling.}
\end{figure}

\begin{figure}[H]
\centering
\includegraphics[width=\textwidth]{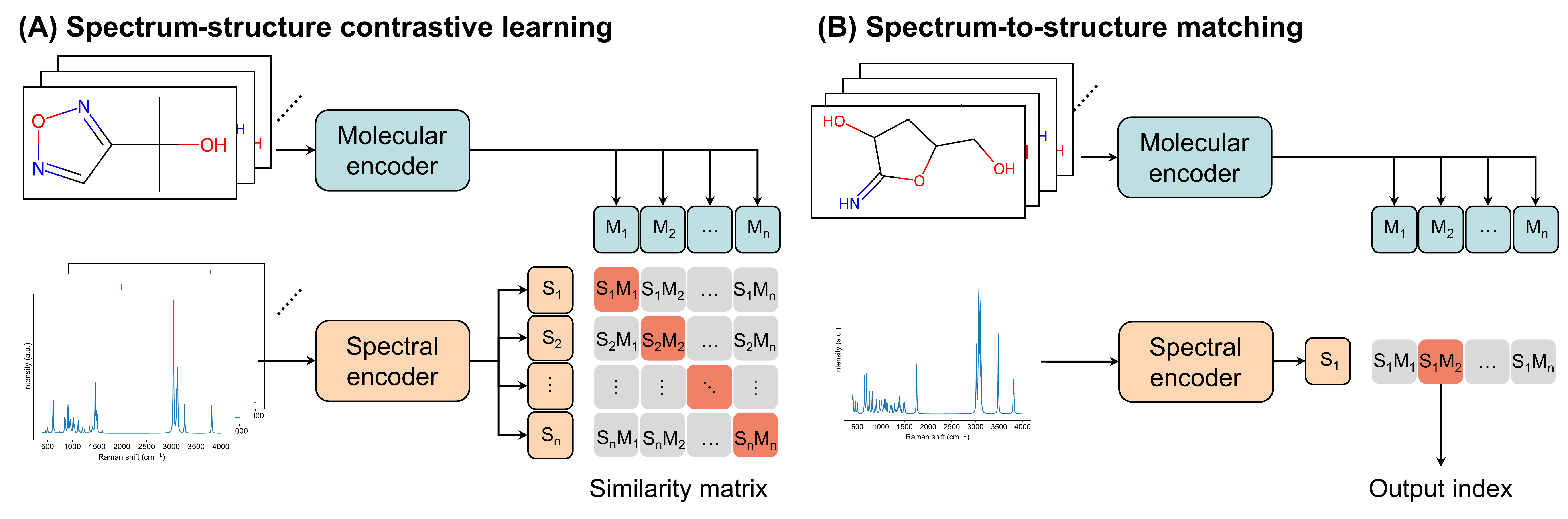}
\caption{Workflow of contrastive learning for (A) training and (B) testing.}
\end{figure}

\begin{figure}[H]
\centering
\includegraphics[width=\textwidth]{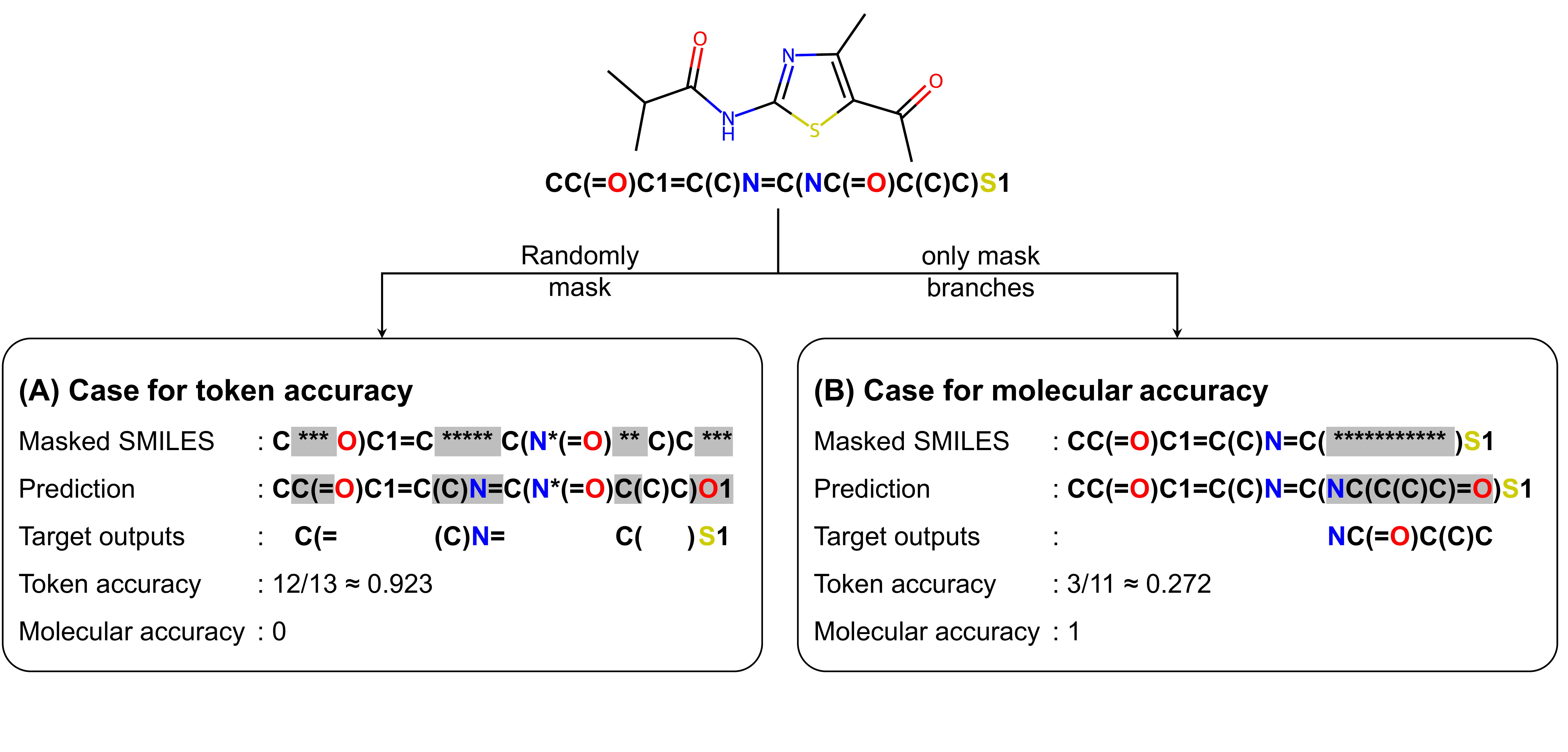}
\caption{Comparison between token accuracy and molecular accuracy.}
\end{figure}

\begin{figure}[H]
\centering
\includegraphics[width=\textwidth]{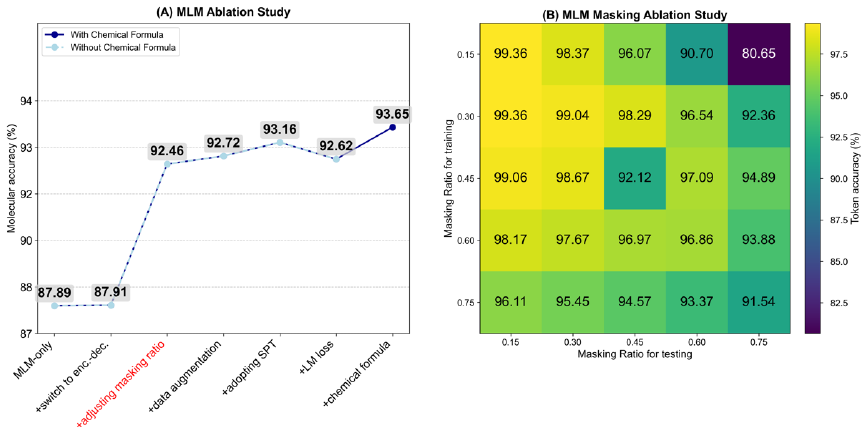}
\caption{Ablation studies of MLM. (A) The factors affecting the performance of MLM evaluated by molecular accuracy. (B) The relationship between the masking ratio used for training and testing evaluated by token accuracy. Models perform best when the training and testing masking ratios are similar. This highlights that a 45\% masking ratio is an optimal hyperparameter, as models trained with this ratio demonstrate robust performance across a wide range of testing conditions.}
\end{figure}

\begin{figure}[H]
\centering
\includegraphics[width=\textwidth]{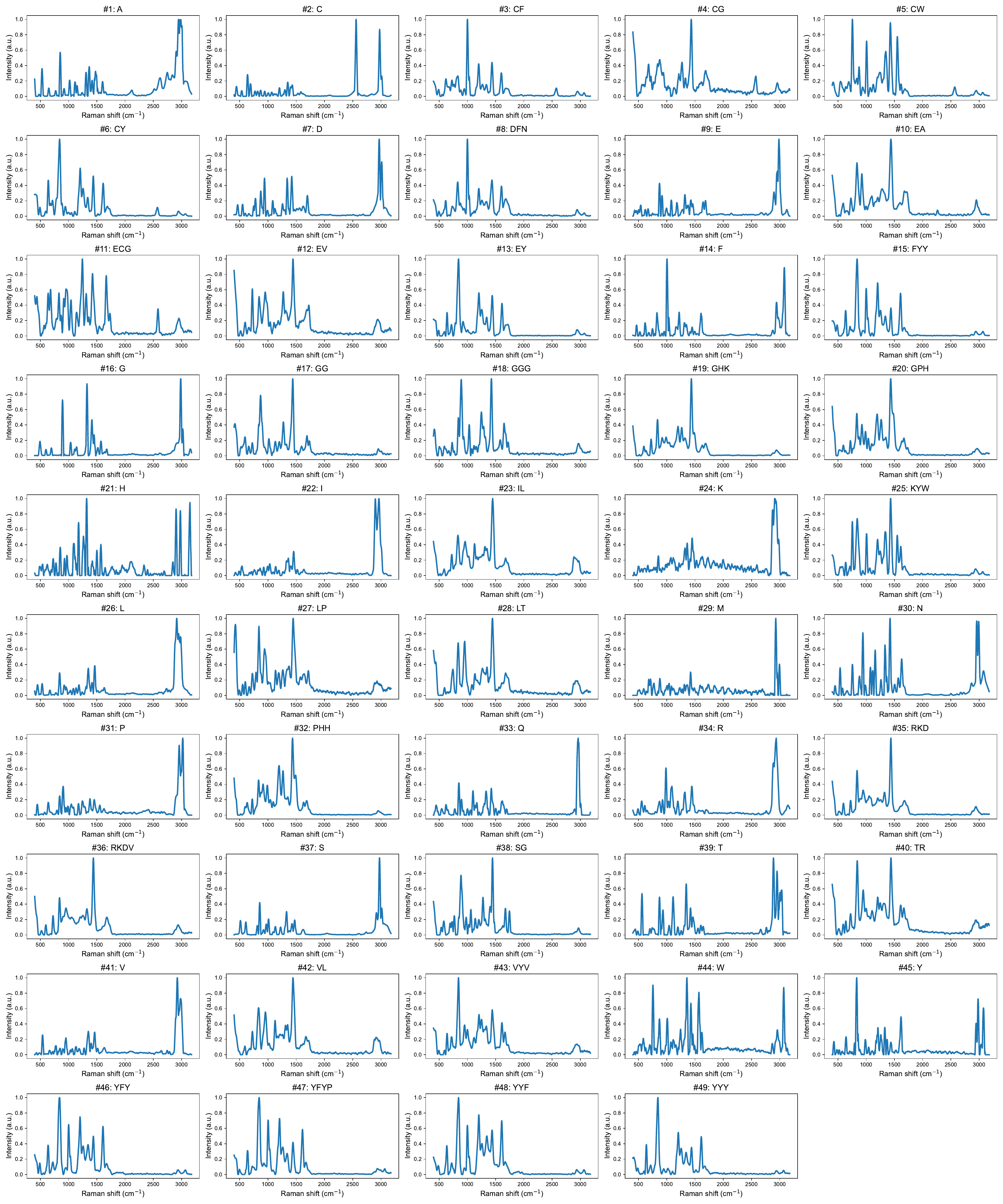}
\caption{Experimental Raman spectra of peptides.}
\end{figure}

% Table generated by Excel2LaTeX from sheet 'Sheet1'
\begin{table}[htbp]
  \centering
  \caption{Add caption}
    \begin{tabular}{p{10em}p{6em}p{6em}p{6em}}
    \toprule
    \multicolumn{1}{c}{} & VibraCLIP & SMEN & Vib2Mol \\
    \midrule
    \multirow{2}[1]{*}{Molecular representation} & atomic numbers and bond features  & atomic numbers and coordinates  & \multirow{2}[1]{*}{SMILES} \\
    \multicolumn{1}{c}{} & (2D graph) & (3D graph) & \multicolumn{1}{c}{} \\
    Molecular encoder & DimeNet++ & EGNN & Transformer \\
    Spectral encoder & MLP & Transformer & Transformer \\
    num of molecular layers & \multicolumn{1}{c}{4} & \multicolumn{1}{c}{5} & \multicolumn{1}{c}{6} \\
    num of spectral layers & \multicolumn{1}{c}{1} & \multicolumn{1}{c}{5} & \multicolumn{1}{c}{6} \\
    Dimension of molecular embedding & \multicolumn{1}{c}{128} & \multicolumn{1}{c}{256} & \multicolumn{1}{c}{768} \\
    Dimension of spectral embedding & \multicolumn{1}{c}{1262} & \multicolumn{1}{c}{512} & \multicolumn{1}{c}{768} \\
    Dropout rate & \multicolumn{1}{c}{0} & \multicolumn{1}{c}{0.1} & \multicolumn{1}{c}{0.1} \\
    Optimizer & AdamW & AdamW & AdamW \\
    Learning rate & \multicolumn{1}{c}{0.0004} & \multicolumn{1}{c}{0.0001} & \multicolumn{1}{c}{0.0001} \\
    Batch size & \multicolumn{1}{c}{128} & \multicolumn{1}{c}{200} & \multicolumn{1}{c}{1024} \\
    Patch size / spectral length & 1 / 1750 & 7 / 1024 & 8 / 1024 \\
    spectral range & 400~4000 & 400~4000 & 400~4000 \\
    Augment types & None & None & None \\
    \bottomrule
    \end{tabular}%
  \label{tab:tables1}%
\end{table}%

% Table generated by Excel2LaTeX from sheet 'Sheet1'
\begin{table}[htbp]
  \centering
  \caption{Hyper-parameters about generation models.}
    \begin{tabular}{p{8em}p{5em}p{5em}p{5em}}
    \toprule
    \multicolumn{1}{c}{} & PBSA & IR2Mol & Vib2Mol \\
    \midrule
    num of heads & \multicolumn{1}{c}{8} & \multicolumn{1}{c}{8} & \multicolumn{1}{c}{8} \\
    num of layers & \multicolumn{1}{c}{4} & \multicolumn{1}{c}{6} & \multicolumn{1}{c}{6} \\
    Embedding dimension & \multicolumn{1}{c}{512} & \multicolumn{1}{c}{512} & \multicolumn{1}{c}{768} \\
    FFN dimension & \multicolumn{1}{c}{2048} & \multicolumn{1}{c}{2048} & \multicolumn{1}{c}{2048} \\
    Dropout rate & \multicolumn{1}{c}{0.1} & \multicolumn{1}{c}{0.1} & \multicolumn{1}{c}{0.1} \\
    Optimizer & Adam & AdamW & AdamW \\
    Learning rate & \multicolumn{1}{c}{0.0001} & \multicolumn{1}{c}{0.001} & \multicolumn{1}{c}{0.0001} \\
    Batch size & \multicolumn{1}{c}{128} & \multicolumn{1}{c}{128} & \multicolumn{1}{c}{128} \\
    Patch size / spectral length & 64 / 3200 & 75 / 1625 & 8 / 1024 \\
    spectral range & 400~3980 & 650~3900 & 400~4000 \\
    Augment types & SMILES & SMILES & SMILES \\
    \bottomrule
    \end{tabular}%
  \label{tab:tables2}%
\end{table}%

% Table generated by Excel2LaTeX from sheet 'Sheet1'
\begin{table}[htbp]
  \centering
  \caption{Spectrum-structure retrieval and de novo generation performance on the QM9S test set, reported as Top-1 Recall (\%).}
    \begin{tabular}{cccccc}
    \toprule
    \multirow{3}[4]{*}{} & \multirow{2}[2]{*}{extra info} & \multicolumn{2}{c}{\multirow{2}[2]{*}{Spectrum-structure retrieval}} & \multicolumn{2}{c}{de novo } \\
      &   & \multicolumn{2}{c}{} & \multicolumn{2}{c}{generation} \\
\cmidrule{2-6}      &   & Raman & IR & Raman & IR \\
    \midrule
    VibraCLIP & None & 97.07 & 95.23 & - & - \\
    PBSA & chemical formula & - & - & 78.77 & 69.21 \\
    IR2Mol & chemical formula & - & - & 80.55 & 75.39 \\
    SMEN & conformation & 97.89 & \textbf{97.04} & 65.78 & 51.54 \\
    Vib2Mol & None & 97.18 & 95 & 89.02 & 82.68 \\
    Vib2Mol & chemical formula & \textbf{98.11} & 96.63 & \textbf{90.91} & \textbf{86.77} \\
    \bottomrule
    \end{tabular}%
  \label{tab:tables3}%
\end{table}%

% Table generated by Excel2LaTeX from sheet 'Sheet1'
\begin{table}[htbp]
  \centering
  \caption{Spectrum-structure retrieval and de novo generation performance on the VB-Mols test set, reported as Top-1 Recall (\%).}
    \begin{tabular}{cccccc}
    \toprule
    \multirow{3}[4]{*}{} & \multirow{2}[2]{*}{extra info} & \multicolumn{2}{c}{Spectrum-structure} & \multicolumn{2}{c}{de novo } \\
      &   & \multicolumn{2}{c}{retrieval} & \multicolumn{2}{c}{generation} \\
\cmidrule{2-6}      &   & Raman & IR & Raman & IR \\
    \midrule
    VibraCLIP & None & 91.21 & 89.01 & - & - \\
    PBSA & chemical formula & - & - & 69.37 & 61.46 \\
    IR2Mol & chemical formula & - & - & 74.79 & 71.44 \\
    SMEN & conformation & \textbf{95.43} & \textbf{94.02} & 53.08 & 42.25 \\
    Vib2Mol & None & 93.2 & 91.38 & 82.59 & 78.59 \\
    Vib2Mol & chemical formula & 94.66 & 93.38 & \textbf{86.74} & \textbf{83.69} \\
    \bottomrule
    \end{tabular}%
  \label{tab:tables4}%
\end{table}%

% Table generated by Excel2LaTeX from sheet 'Sheet1'
\begin{table}[htbp]
  \centering
  \caption{Spectrum-spectrum retrieval performance on the VB-GEOM test set, reported as Top-1 Recall (\%).}
    \begin{tabular}{ccc}
    \toprule
    \multirow{3}[4]{*}{} & \multicolumn{2}{c}{Spectrum-spectrum} \\
      & \multicolumn{2}{c}{retrieval} \\
\cmidrule{2-3}      & Raman & IR \\
    \midrule
    Cosine similarity & 36.62 & 34.28 \\
    Pearson correlation coefficient & 35.86 & 33.39 \\
    VibraCLIP & 75.22 & 70.16 \\
    SMEN & 41.8 & 49.01 \\
    Vib2Mol & \textbf{77.54} & \textbf{75.33} \\
    \bottomrule
    \end{tabular}%
  \label{tab:tables5}%
\end{table}%

% Table generated by Excel2LaTeX from sheet 'Sheet1'
\begin{table}[htbp]
  \centering
  \caption{Spectrum-structure retrieval and de novo generation performance on the SDBS test set, reported as Top-1 Recall (\%)}
    \begin{tabular}{cccccc}
    \toprule
    \multirow{3}[4]{*}{} & \multirow{3}[4]{*}{extra info} & \multicolumn{2}{c}{Spectrum-structure} & \multicolumn{2}{c}{de novo } \\
      &   & \multicolumn{2}{c}{retrieval} & \multicolumn{2}{c}{generation} \\
\cmidrule{3-6}      &   & Raman & IR & Raman & IR \\
    \midrule
    VibraCLIP & None & 68.57 & 74.28 & - & - \\
    PBSA & chemical formula & - & - & 41.29 & 30.68 \\
    IR2Mol & chemical formula & - & - & 50.71 & 48.22 \\
    Vib2Mol & None & 78.01 & 71.28 & 24.11 & 24.82 \\
    Vib2Mol & chemical formula & \textbf{90.43} & \textbf{86.17} & \textbf{56.03} & \textbf{52.84} \\
    \bottomrule
    \end{tabular}%
  \label{tab:tables6}%
\end{table}%

% Table generated by Excel2LaTeX from sheet 'Sheet1'
\begin{table}[htbp]
  \centering
  \caption{Spectrum-structure retrieval and de novo generation performance on the NIST-IR test set, reported as Top-1 Recall (\%).}
    \begin{tabular}{cccc}
    \toprule
    \multirow{2}[2]{*}{} & \multirow{2}[2]{*}{extra info} & Spectrum-structure  & \multicolumn{1}{p{5em}}{de novo } \\
      &   & retrieval & \multicolumn{1}{p{5em}}{generation} \\
    \midrule
    VibraCLIP & None & 67.47 & \multicolumn{1}{p{5em}}{-} \\
    PBSA & chemical formula & - & 41.56 \\
    IR2Mol & chemical formula & - & 50.78 \\
    Vib2Mol & None & 70.78 & 34.07 \\
    Vib2Mol & chemical formula & \textbf{83.54} & \textbf{55.14} \\
    \bottomrule
    \end{tabular}%
  \label{tab:tables7}%
\end{table}%

% Table generated by Excel2LaTeX from sheet 'Sheet1'
\begin{table}[htbp]
  \centering
  \caption{Large-scale retrieval performance of Vib2Mol-MM against the PubChem database.}
    \begin{tabular}{p{15em}p{10em}}
    \toprule
    Additional molecules from PubChem & Recall@1 / Recall@5 \\
    \midrule
    0 (only original test set) & 95.74 / 97.87 \\
    \multicolumn{1}{c}{1,000} & 95.39 / 96.45 \\
    \multicolumn{1}{c}{10,000} & 89.36 / 90.43 \\
    \multicolumn{1}{c}{100,000} & 83.69 / 84.40 \\
    \multicolumn{1}{c}{1,000,000} & 68.44 / 69.86 \\
    \multicolumn{1}{c}{10,000,000} & 48.23 / 51.42 \\
    80,368,968 (total PubChem) & 30.60 / 33.81 \\
    \bottomrule
    \end{tabular}%
  \label{tab:tables8}%
\end{table}%

% Table generated by Excel2LaTeX from sheet 'Sheet1'
\begin{table}[htbp]
  \centering
  \caption{Latent space similarity analysis of experimental peptides. Values represent mean cosine similarity (\%) between a sample’s embedding and the rest of the dataset.}
    \begin{tabular}{p{5em}p{15em}p{15em}}
    \toprule
    \multicolumn{1}{c}{} & Structural similarity & Spectral similarity \\
    \midrule
    Dipeptide & 55.74±19.91 & 50.34±21.23 \\
    Tripeptide & 51.69±24.04 & 49.45±22.92 \\
    Tetrapeptide & 51.51±24.04 & 49.10±22.92 \\
    Averaged & 53.64±22.19 & 49.86±22.27 \\
    \bottomrule
    \end{tabular}%
  \label{tab:tables9}%
\end{table}%

% Table generated by Excel2LaTeX from sheet 'Sheet1'
\begin{table}[htbp]
  \centering
  \caption{Details about the data split for training, validation, and testing.}
    \begin{tabular}{p{5em}p{5em}p{5em}ccc}
    \toprule
    Datasets & Data source & Modality & \multicolumn{1}{p{5em}}{Training size} & \multicolumn{1}{p{5em}}{Evaluating size} & \multicolumn{1}{p{5em}}{Testing size} \\
    \midrule
    QM9S & Theoretical & IR+Raman & 110992 & 5842 & 12982 \\
    VB-QM9 & Theoretical & IR+Raman & 93403 & 13344 & 26687 \\
    VB-ZINC15 & Theoretical & IR+Raman & 38089 & 5442 & 10883 \\
    VB-mols & Theoretical & IR+Raman & 131492 & 18786 & 37570 \\
    VB-GEOM & Theoretical & IR+Raman & 0 & 0 & \multicolumn{1}{p{5em}}{6835*2} \\
    \midrule
    SDBS & Experimental & IR+Raman & 2393 & 140 & 282 \\
    NIST-IR & Experimental & IR & 10327 & 602 & 1215 \\
    \midrule
    PAHs & Theoretical & Raman & 3006 & 430 & 860 \\
    RXN & Theoretical & Raman & 10947 & 1564 & 3128 \\
    \midrule
    Peptide & Theoretical & Raman & 18168 & 2596 & 5191 \\
    Peptide-mod & Theoretical & Raman & 8787 & 1256 & 2511 \\
    \bottomrule
    \end{tabular}%
  \label{tab:tables10}%
\end{table}%

% Table generated by Excel2LaTeX from sheet 'Sheet1'
\begin{table}[htbp]
  \centering
  \caption{Comparison of model size, computational complexity, and memory requirements.}
    \begin{tabular}{p{5em}p{5em}cccc}
    \toprule
    \multicolumn{1}{r}{} & Training phase & \multicolumn{1}{p{5em}}{Computational cost (GFLOPS)} & \multicolumn{1}{p{5em}}{\# of parameters (M)} & \multicolumn{1}{p{5em}}{\# of active parameters (M)} & \multicolumn{1}{p{5em}}{Memory requirements (G/GPU)} \\
    \midrule
    CL-only & alignment & 13.2 & 66.66 & 66.66 & 36.87 \\
    CL+Matching & alignment & 35.02 & 114.1 & 114.1 & 75.27 \\
    Vib2Mol & alignment & 35.02 & 145.86 & 81.03 & 69.03 \\
    \midrule
    LM-only & generation & 15.95 & 80.48 & 80.48 & 5.42 \\
    MLM-only & generation & 16.04 & 80.48 & 80.48 & 5.36 \\
    LM+MLM & generation & 23.16 & 128.36 & 128.36 & 7.96 \\
    Vib2Mol & generation & 27.47 & 145.86 & 113.55 & 9.08 \\
    \bottomrule
    \end{tabular}%
  \label{tab:tables11}%
\end{table}%

%%=============================================%%
%% For submissions to Nature Portfolio Journals %%
%% please use the heading ``Extended Data''.   %%
%%=============================================%%

%%=============================================================%%
%% Sample for another appendix section			       %%
%%=============================================================%%

%% \section{Example of another appendix section}\label{secA2}%
%% Appendices may be used for helpful, supporting or essential material that would otherwise 
%% clutter, break up or be distracting to the text. Appendices can consist of sections, figures, 
%% tables and equations etc.

\end{appendices}

%%===========================================================================================%%
%% If you are submitting to one of the Nature Portfolio journals, using the eJP submission   %%
%% system, please include the references within the manuscript file itself. You may do this  %%
%% by copying the reference list from your .bbl file, paste it into the main manuscript .tex %%
%% file, and delete the associated \verb+\bibliography+ commands.                            %%
%%===========================================================================================%%
\clearpage
\bibliography{sn-bibliography}% common bib file
%% if required, the content of .bbl file can be included here once bbl is generated
%%\input sn-article.bbl

\end{document}